\documentclass[aps,prb,superscriptaddress,twocolumn,floatfix,showpacs,citeautoscript,longbibliography,hyperlinks]{revtex4-2}
\usepackage{amsmath}
\usepackage{amssymb}
\usepackage{amsfonts}
\usepackage{dsfont}
\usepackage{color}
\usepackage{nicefrac}
\usepackage[autostyle]{csquotes}
\usepackage[colorlinks=true,citecolor=blue]{hyperref}
\usepackage{graphicx,graphics}
\usepackage{siunitx}

\usepackage{soul}

\newcommand{\revision}[1]{{{#1}}}

\begin{document}
\title{Heat transport in a two-level system driven by a time-dependent temperature}

\author{Pedro Portugal}
\affiliation{Department of Applied Physics, Aalto University, 00076 Aalto, Finland}
\author{Christian Flindt}
\affiliation{Department of Applied Physics, Aalto University, 00076 Aalto, Finland}
\author{Nicola Lo Gullo}
\affiliation{CSC–IT Center for Science, P.O.~Box 405, 02101 Espoo, Finland}

\begin{abstract}
The field of thermotronics aims to develop thermal circuits that operate with temperature biases and heat currents just as how electronic circuits are based on voltages and electric currents. Here, we investigate a thermal half-wave rectifier based on a quantum two-level system (a qubit) that is driven by a periodically modulated temperature difference across it. To this end, we present a non-equilibrium Green's function technique, which we extend to the time domain to account for the time-dependent temperature in one of two thermal reservoirs connected to the qubit. We find that the qubit acts a thermal diode in parallel with a thermal capacitor, whose capacitance is controlled by the coupling to the reservoirs. These findings are important for the efforts to design non-linear thermal components such as heat rectifiers and multipliers that operate with more than one diode.
\end{abstract}

\maketitle

\section{Introduction}
The miniaturization of electronic elements has led to a dramatic growth in the density of components in electronic circuits \cite{Streetman2016}. This remarkable progress has allowed the design of more compact and faster electronic devices, and it has increased the computational throughput of the individual processing units. At the same time, the dissipation of heat at micro and nano scales has become a serious issue, which gets increasingly adverse as electronic circuits are scaled down in size. In one attempt to avoid overheating, the waste heat may be dissipated into the environment by means of a refrigeration circuit \cite{Giazotto2006,Pekola2015}. However, this approach may greatly reduce the overall efficiency of a device, not only because of the wasted energy, but also because the refrigeration circuit itself requires external power, which can be a significant fraction of the total power consumption of a device~\cite{Jones2018}.

In an alternative way to handle waste heat generated by electronic devices, one may try to exploit thermoelectric effects and thereby convert and store parts of the energy for use at a later time \cite{Shakouri2011,Biehs2017}. While this strategy might be an improvement over simply dissipating the waste heat into the environment, it is still rather inefficient \cite{vining2009inconvenient}, and more radical ideas may be needed to exploit waste heat. In the approach that we follow here, the aim is to build heat-flow based components, which make use of heat currents and temperature gradients just as how electric currents and voltages are controlled in electronic circuits. The management of heat carried by phonons is known as phononics \cite{Dhar2008,hanggi2012}, and the more general field dealing with heat flows in systems involving electrons~\cite{Wu2009}, photons \cite{Pascal2011,Ben-Abdallah2011,Brange2019}, or phonons has been coined thermotronics~\cite{Biehs2017}.

\begin{figure}
	\includegraphics[width=0.95\columnwidth]{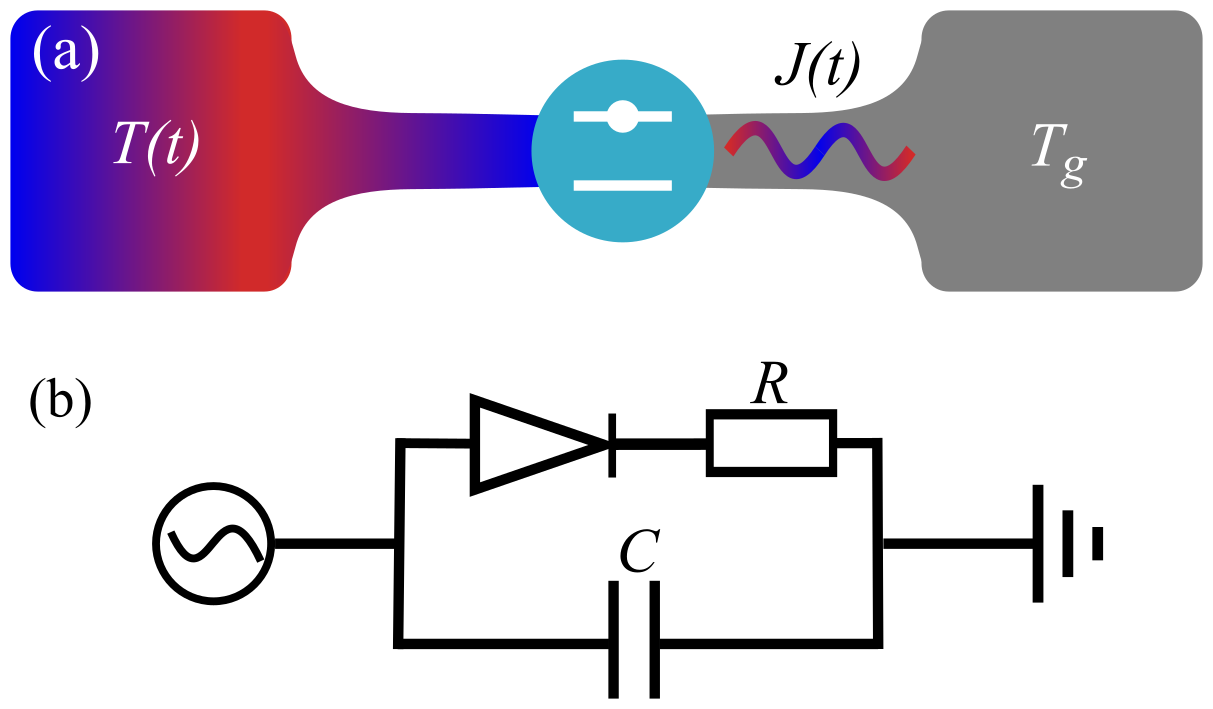}
	\centering
	\caption{AC-temperature driven qubit. (a) The setup consists of a quantum two-level system, or qubit, coupled to two thermal reservoirs. A time-dependent temperature in the left reservoir, $T(t)$, generates a time-dependent heat current, $J(t)$, in the right reservoir, which is kept at the constant base temperature, $T_g$. (b) Equivalent circuit diagram of the quantum two-level system, which acts as a thermal diode in series with a resistor, $R$, and in parallel with a capacitor, $C$. }\label{fig:scheme}
\end{figure}

Although the field of thermotronics is still in its infancy, several basic components have already been proposed and realized such as thermal transistors \cite{Ojanen2008,Ruokola2009}, thermal memories \cite{Peotta2014,Kubytsky2014,Ordonez2019}, thermal heat valves \cite{Ronzani2018,LoGullo2020}, and thermal switches \cite{Dutta2017}. When combined, these elements may pave the way for logic circuits that process a thermal input signal and return an output signal in terms of a heat current \cite{wang2007thermal}. One key component for such circuits is a thermal diode, which allows for non-reciprocal transport of heat currents upon the inversion of the thermal gradient. A simple and promising realization of thermal diodes is based on spin-boson systems, where a quantum two-level system is coupled to two thermal baths at different temperatures \cite{Segal2005,Carrega2015}. Such systems are known to exhibit non-reciprocal heat transport as needed for a thermal diode \cite{velizhanin2010meir,PhysRevA.95.023610,Yang2014,Senior2020,belyansky2020transport}, and they have recently been used to realize a thermal heat valve~\cite{Ronzani2018}. \revision{While earlier works have considered low-frequency heat transport~\cite{Eich_2016,PhysRevB.90.115116,PhysRevLett.114.196601,PhysRevB.102.155407}, we now extend the discussion to finite driving frequencies.}

In the framework of thermotronics, we here theoretically investigate the application of a thermal diode as a half-wave thermal rectifier. We consider the setup in Fig.~\ref{fig:scheme}a consisting of a quantum two-level system (or qubit) coupled to two bosonic baths. The right reservoir it kept at a constant temperature, while the left one is driven by a time-dependent temperature, encoding a thermal input signal. The output signal is then given by the time-dependent heat current flowing into the right reservoir.
To evaluate the time-dependent heat current, we extend the method of Refs.~\cite{liu2016green,liu2017} for static systems to setups with  time-dependent temperatures. We then apply this methodology to characterize the qubit as a thermal half-wave rectifier. Based on the response to a dynamic temperature bias, we find that the system exhibits a non-trivial dependence on both the driving frequency and amplitude. \revision{Our approach is valid at slow driving frequencies compared to the qubit spacing and with not too strong couplings. We find that} the system behaves as a diode in parallel with a thermal capacitor as illustrated in Fig.~\ref{fig:scheme}b. The capacitance induces a phase shift between the input and output signals, and we show that its microscopic origin can be related to the coupling between the two-level system and the heat baths.

Our work is organized as follows. In Sec.~\ref{sec:qubit}, we introduce the transport setup consisting of a quantum two-level system coupled to two heat baths, and we discuss the implementation of a time-dependent temperature. In Sec.~\ref{sec:NEGF}, we describe the non-equilibrium Green's function (NEGF) approach that we use to calculate the time-dependent heat current running into the drain reservoir. In Sec.~\ref{sec:DC}, we consider a constant temperature difference between the baths to investigate the low-frequency transport properties of the quantum two-level system, before turning to the general case of a time-dependent temperature in Sec.~\ref{sec:AC}. There, we evaluate the time-dependent heat current due to a periodic temperature drive, and we develop an equivalent circuit model for the quantum two-level system, which acts as a thermal diode in series with a small resistance and in parallel with a capacitor. We also provide simple estimates of the system parameters, which are relevant for realistic physical setups. Finally, in Sec.~\ref{sec:concl}, we summarize our findings and provide an outlook on possible directions for future work.

\section{Thermal circuit with a qubit}
\label{sec:qubit}

Our goal is to characterize the dynamic thermal transport properties of a qubit in a two-terminal setup as depicted in Fig.~\ref{fig:scheme}a. In the weak coupling regime without a time-dependent drive, the system is known to behave as a thermal diode~\cite{Segal2005}, which led to the recent realization of a thermal heat-valve~\cite{Ronzani2018}. Here, our aim is investigate the setup with a time-dependent temperature in the spirit of thermotronics~\cite{Biehs2017}. We will drive the temperature of one bath with frequency $\Omega$ and amplitude $\Delta T$ and evaluate the heat current running into the other bath.

The setup consists of a quantum two-level system (a qubit) coupled to left and right heat reservoirs. In this work, we focus on two heat baths, but our approach can readily be extended to more complex setups such as a three-terminal system making up a transistor.
The system is described by the Hamiltonian
\begin{equation}
\hat{H}(t)=\hat{H}_{\rm Q}+\sum\limits_{\ell=L,R
}\left[\hat{H}_\ell
(t)+\hat{T}_{\ell
}(t)\right],
\end{equation}
where the qubit is given by the term
\begin{equation}
\hat{H}_{\rm Q}=\frac{\hbar\varepsilon}{2}\hat{\sigma}_x,
\end{equation}
while the Hamiltonian of each heat bath takes the form
\begin{equation}
\hat{H}_{\ell}(t)= r_\ell(t)\sum\limits_{k}\hbar\omega_{\ell k} \hat a_{\ell k}^\dagger \hat a_{\ell k},
\end{equation}
where $\hat a_{\ell k}^\dagger$ and $\hat a_{\ell k}$ describe bosonic modes of frequency $\omega_{\ell k}$, and we take $\hbar, k_B=1$ from now on. The terms
\begin{equation}
\hat T_\ell(t)=r_\ell(t)\sum\limits_{k}s_\ell(t,\omega_{\ell k}) g_{\ell k} \hat \sigma_z \hat q_{\ell k}
\label{eq:intH}
\end{equation}
describe the coupling between the qubit and the heat baths, where $g_{\ell k}$ is the strength of the coupling to a bath excitation with momentum $k$, and we have defined the position operator $\hat q_{\ell k}=\hat a_{\ell k}^\dagger+ \hat a_{\ell k}$. Above, we have also introduced the time-dependent parameters $r_\ell(t)$ and $s_\ell(t,\omega_{\ell k})$, which will be important in the following for the description of time-dependent temperatures.

\subsection{Time-dependent temperatures}

To describe a time-dependent temperature we exploit an analogy with time-dependent voltages in electronic setups. In that case, a modulated voltage is described by a time-dependent local chemical potential in each bath. The central idea is that the chemical potential is the Lagrange multiplier that guarantees that the mean number of particles in the reservoir is kept constant.
Similarly, the inverse temperature is the Lagrange multiplier that fixes the mean energy of the external reservoir. Hence, the time-dependent scaling factor $r_\ell(t)$ in $\hat{H}_\ell$ plays the role of the multiplier $\mu$ (the chemical potential) in front of the operator for the total particle number in the grand canonical ensemble. Thus, it scales the energies of the bath and, as we will see, it effectively leads to a rescaling of the inverse bath temperature, which becomes $\beta_\ell /r_\ell(t)$. 

A related approach based on a functional theory~\cite{Eich_2016} has been presented in \revision{Refs.~\cite{PhysRevB.90.115116,PhysRevLett.114.196601,PhysRevB.102.155407} using a Luttiger-field method \cite{PhysRev.135.A1505},} however, without the rescaling of the interaction term  $\hat T_\ell(t)$ that we include here. Specifically, the prefactor $r_\ell(t)s_\ell(t,\omega_{\ell k})$ in the coupling Hamiltonian is chosen so that the rescaling of the energies does not change the spectral properties of the bath. In particular, when rescaling the bath Hamiltonian, the density of states is modified, and the frequency of a particular mode is changed from $\omega_{\ell k}$ to $r_\ell(t)\omega_{\ell k}$. As a result, the spectral function, which is given by the couplings $g_{\ell k}$, is effectively changed. We thus include the prefactor $s_\ell(t,\omega_{\ell k})$ in front of the coupling to keep the spectral density unchanged. The specific form of $s_\ell(t,\omega_{\ell k})$ will depend on the couplings $g_{\ell k}$ and will be discussed in Sec.~\ref{sec:NEGF}.  Below, we only drive the temperature of the left bath, and we thus keep $r_R(t)=1$ and $s_R(t,\omega_{\ell k})=1$ for the right bath.

\subsection{Time-dependent heat current}
To investigate the thermal transport properties of the qubit, we will consider how it transforms a periodic input signal, encoded in the temperature drive of the left bath, into an output signal in the right bath. We express the time-dependent heat current in the right bath as~\cite{esposito2015quantum,velizhanin2010meir} 
\begin{equation}
J_R(t)=-i\left\langle\left[\hat{H}(t),\hat{H}_R\right]\right\rangle,
\end{equation}
where the Hamiltonian of the bath, $\hat{H}_R$, is time-independent, since the temperature there is kept constant. Here, we consider the part of the heat current that is given by the rate of change of the bath energy~\cite{PhysRevB.92.235440}. \revision{On the other hand, we do not discuss the one that is associated with the contact region between the system and the heat bath, but note that this term may need to be considered to establish a thermodynamically consistent framework at finite frequencies~\cite{Ludovico2014,Ludovico2016a,Ludovico2016b,Ludovico2018}.} The time-dependent heat current can  now be evaluated using non-equilibrium Green's functions, and following the derivation in App.~\ref{app:hcur}, it can be written as
\begin{equation}
J_R(t)=2{\rm Re}\left[\int d\omega \omega I_R(\omega) \mathcal{J}_R(\omega)\right]
\label{eq:hcurdef}
\end{equation}
where 
\begin{equation}
\mathcal{J}_R(\omega)=\int d\tau \left[\mathcal{D}_{R\omega}^R(t,\tau)K_z^<(\tau,t)+\mathcal{D}_{R\omega}^<(t,\tau)K_z^A(\tau,t)\right]
\end{equation}
is the average flow of bosons into the right reservoir at a given frequency, and $I_R(\omega)$ is the spectral density of that bath. To evaluate the heat current, we then need to compute the lesser two-times self-correlation function 
\begin{equation}
K_z^<(t,t')=-i\left\langle\hat{\sigma}_z(t')\hat{\sigma}_z(t)\right\rangle
\end{equation}
of the qubit as well as its spectral function encoded in the advanced component 
\begin{equation}
K_z^A(t,t')=\theta(t-t')(K_z^<(t,t')-K_z^>(t,t')),
\end{equation}
where $\theta$ is the Heaviside step function. Moreover we need the free propagators of each bath mode, $\mathcal{D}_{\ell\omega}^{R,<}(t,t')$.

\section{NEGF for a driven system}
\label{sec:NEGF}
To evaluate the heat current, we use non-equilibrium Green's functions to find the time-dependent correlation functions of the qubit. In particular, we extend the approach of Refs.~\cite{liu2017,liu2016green} to include a time-dependent temperature. 
We start by performing a polaron transformation, which allows us to rewrite the coupling to the reservoirs in a manner that can be treated using perturbation theory~\cite{xu2016polaron,hsieh2019nonequilibrium,xu2016non}. Next, we represent the spin operators using Majorana fermions~\cite{PhysRevLett.91.207203,PhysRevB.93.174420}. The commutation relations of the spin operators make it difficult to apply Wick's theorem, but we may switch to a Majorana fermion representation, which has proven useful to evaluate spin-spin correlation functions perturbatively~\cite{liu2017,agarwalla2017,xu2016non}. In the following, we only describe the main steps of this calculation and refer the reader to App.~\ref{app:dyson} for further details. Our approach follows Ref.~\cite{liu2017}, but for the sake of completeness and to highlight the additional steps that are required for the time-dependent drive, we present the essential details here.

\subsection{Polaron transformation}
We first apply a polaron transformation of the form
\begin{equation}
\hat{U}(t)=e^{-i\hat{\sigma}_z \hat{\Omega}(t)},
\end{equation}
where we have defined the operator
\begin{equation}
\hat{\Omega}(t)=2 i\sum\limits_\ell\sum\limits_k s_\ell(t,\omega_{\ell k}) g_{\ell k}\omega_{\ell k}^{-1}(\hat{a}_{\ell k}^\dagger-\hat{a}_{\ell k}).
\label{eq:Omega}
\end{equation}
The unitary operator $\hat{U}(t)$ shifts the equilibrium position of the bath oscillators according to the state of the qubit. Specifically, in the polaron frame, \revision{the Hamiltonian reads
\begin{equation}
	\begin{split}
		\tilde{H}(t)&=\hat{U}(t)\hat{H}(t)\hat{U}^\dagger(t)+i\hbar \left[\frac{d}{dt}\hat U(t)\right]\hat U^\dagger(t)\\
		&=\check{H}(t)+\sum_\ell\hat{H}_\ell(t)+\hat H _U (t),
	\end{split}
\end{equation}
where the first term
\begin{equation}
\check{H}(t)=\frac{\displaystyle\varepsilon}{\displaystyle 2}\left(\hat{\sigma}_x\cos\hat{\Omega}(t)+
\hat{\sigma}_y\sin\hat{\Omega}(t)\right),
\end{equation}
contains the resummed interactions between the qubit and the baths, which are bounded by $\varepsilon$ because of the trigonometric functions. The last term above
\begin{equation}
\hat H_U(t) =2 i\hbar\hat\sigma_z \sum\limits_\ell\sum\limits_k \frac{d s_\ell(t,\omega_{\ell k})}{dt} g_{\ell k}\omega_{\ell k}^{-1}(\hat{a}_{\ell k}^\dagger-\hat{a}_{\ell k}),
\end{equation}
is proportional both to the coupling strength and the driving frequency via the time derivative. Thus, below, when we consider weak couplings and driving frequencies that are smaller than the qubit spacing, we can safely neglect this term (and it vanished if the rescaling is constant). We note that the polaron transformation is useful for evaluating multi-time correlators of $\hat{\sigma}_z$, but less so for other correlation functions.}

\subsection{Majorana-fermion representation}
It is not straightforward to apply the NEGF approach to systems described by spin operators such as the Pauli matrices because of their non-trivial commutator algebra. In particular, since neither their commutators nor anti-commutators are just complex numbers, we cannot apply Wick's theorem to reduce higher-order correlators to products of two-point correlation functions. To circumvent this problem, it is convenient to resort to mappings to either fermionic or bosonic operators, depending on the problem at hand. In our case, it is useful either to map the two-level system to a pair of spinless Dirac fermions or to three Majorana fermions~\cite{PhysRevLett.91.207203}. Here, we choose the second strategy and map the spin operators to fictitious Majorana fermions as
\begin{equation}
\hat {\sigma}_{k}=-\frac{i}{2} \epsilon_{klm} \hat{\eta}_{l} \hat{\eta}_{m},
\end{equation} 
where $\epsilon_{klm}$ is the fully anti-symmetric Levi-Civita tensor, summation over repeated indices is implied, and the operators $\hat{\eta}_k$ fulfill the standard Majorana anti-commutation relations, $\{\hat{\eta}_{k},\hat{\eta}_{l}\}=2\varepsilon_{kl}$. We can now introduce the Green's function in the Majorana representation as 
\begin{equation}
G_k(\tau,\tau')=-i\langle \mathcal{T}_\gamma\hat{\eta}_{k}(\tau)\hat{\eta}_{k}(\tau')\rangle,\quad k=x,y,z,
\end{equation}
where $\tau$ and $\tau'$ denote complex times on the Keldysh-Schwinger contour $\gamma$ with the time-ordering operator~$\mathcal{T}_\gamma$~\cite{keldysh1965diagram}. The free Green's functions are  
\begin{equation}
G_{k0}(\tau,\tau')=-i\langle \mathcal{T}_\gamma\hat{\eta}_{k0}(\tau)\hat{\eta}_{k0}(\tau')\rangle,
\end{equation}
where $\hat{\eta}_{k0}(\tau)$ is the Majorana operator in the Heisenberg picture of the uncoupled qubit. One can establish a relationship between the Green's functions in the two pictures \revision{using the expressions below following Ref.~\cite{agarwalla2017}} 
\begin{equation}
\begin{split}
K_{z}^<(t,t')&=-G_{z}^<(t,t'),\\
K_{z}^>(t,t')&=G_{z}^>(t,t'),\\
K_{z}^A(t,t')&=-\Theta(t'-t) [G_{z}^>(t,t')+G_{z}^<(t,t')],\\
K_{z}^R(t,t')&=\Theta(t-t') [G_{z}^>(t,t')+G_{z}^<(t,t')].
\end{split}
\end{equation}
From the expression of the heat current in Eq.~(\ref{eq:hcurdef}), we see that we need to compute the Green's function $G_z(t,t')$.

\subsection{Dyson equation}
To find the Green's function, we first rewrite the Hamiltonian with the resummed interactions as
\begin{equation}
\check{H}(t)=-i\frac{\displaystyle\varepsilon}{\displaystyle 2}\left(\hat{\eta}_y\hat{\eta}_z\cos\hat{\Omega}(t)+
\hat{\eta}_z\hat{\eta}_x\sin\hat{\Omega}(t)\right).
\label{eq:MFham}
\end{equation}
As in Ref.~\cite{liu2017} we then write down a Dyson equation for the full Green's function of the form
\begin{align}
G_z(\tau,\tau')\!=\!G_{z0}(\tau,\tau')\!+\!\!\int\limits_\gamma \!\mathrm{d}s\mathrm{d}s' G_{z0}(\tau,s')\Sigma(s',s)G_z(s,\tau'),
\label{eq:MFDyson}
\end{align}
where the self-energy is expressed as
\begin{equation}
\Sigma(\tau,\tau')=\frac{i}{4}\left[G_{x0}(\tau,\tau') B_{x}(\tau,\tau')+G_{y0}(\tau,\tau') B_{y}(\tau,\tau')\right]
\label{eq:MFSE}
\end{equation}
in terms of the additional bath correlation functions
\begin{equation}
\begin{split}
B_{x}(\tau,\tau')&=-i\left\langle \mathcal{T}_\gamma\cos\hat{\Omega}(\tau)\cos\hat{\Omega}(\tau')\right\rangle,\\
B_{y}(\tau,\tau')&=-i\left\langle \mathcal{T}_\gamma\sin\hat{\Omega}(\tau)\sin\hat{\Omega}(\tau')\right\rangle.
\end{split}
\end{equation}
Due to the polaron transformation, the self-energy is non-additive in the baths, which is different from a master equation, where the contributions from different baths are typically additive.
 
\begin{figure*}
	\centering
	\includegraphics[width=0.98\textwidth]{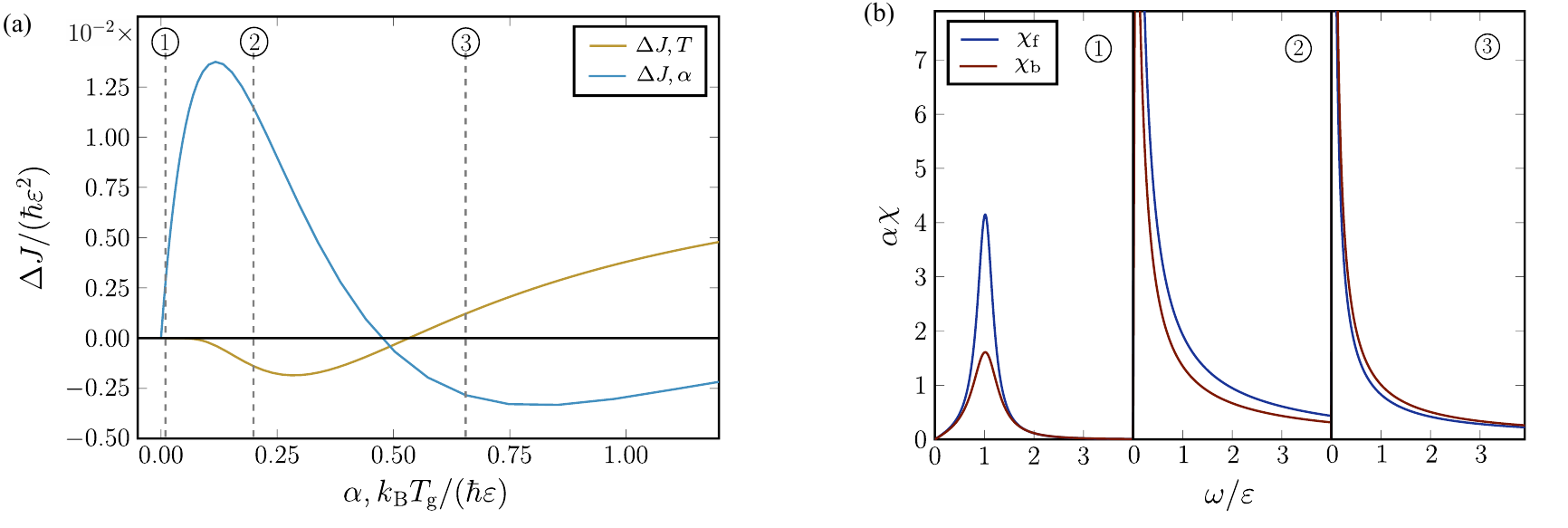}\hfill
	\caption{Rectification and susceptibilities. (a) Difference between the forward and the backward current as a function of the coupling strength (blue) and the ground temperature (yellow) with $k_BT_g=2.5\hbar\varepsilon$ and $\Delta T=0.5 T_g$ (blue) and  $\alpha=0.05$ (yellow). (b) Susceptibilities for the forward (blue) and backward (red) currents, corresponding to the couplings indicated in panel (a).}
	\label{fig:diode}
\end{figure*}

In order to evaluate the baths correlation functions we need to specify the spectral functions for each of them. If we consider spectral functions of the form, 
\begin{equation}
I_\ell(\omega)=\alpha_\ell\pi \omega^p \omega_c^{1-p} e^{-\omega/\omega_c},
\label{eq:bathspectra}
\end{equation}
it is possible to evaluate the self-energies analytically, see App.~\ref{app:dyson}. Here, the dimensionless parameter $\alpha$ controls the strength of the interactions, and $\omega_c$ is a cut-off frequency for the bath. In the ohmic case, $p=1$, the  lesser and greater self energies reads 
\begin{equation}
\Sigma^>(t,t')=[\Sigma^<(t,t')]^*= - i \varepsilon\prod_{\ell=L,R} \Phi^{(\ell)}(t,t'),
\end{equation}
having defined the function
\begin{equation}
\Phi^{(\ell)}(t,t')= \left(\frac{ \omega _c^2 \Gamma^2 [T_\ell(t)/\omega_c+1] \Gamma^2
   [T_\ell(t')/\omega_c+1]}{ T_\ell(t)T_\ell(t')\Gamma^2 [c^{(\ell)}_+(t,t')] \Gamma^2[c^{(\ell)}_-(t,t')+1]}\right)^{\alpha_\ell}.
   \label{eq:Phi-func}
\end{equation}
Above, the gamma function is denoted by $\Gamma[x]$, and 
\begin{equation}
c^{(\ell)}_\pm(t,t^\prime)=[T_\ell(t)+T_\ell(t')]/2\omega_c \pm i \int_{t'}^t \mathrm{d} q T_\ell(q)
\label{eq:c-func}
\end{equation}
contains the effective temperature $T_\ell(t)\equiv r_\ell(t) T_\ell$. Here, for $p=1$, we have made use of the relation
\begin{equation}
s_{\ell}(t,\omega)=[r_{\ell}(t)]^{\frac{p-2}{2}}e^{ \frac{1-r_{\ell}(t)}{2}\omega/\omega_c},
\end{equation}
which we derive in App.~\ref{app:dyson}. We note that with a constant temperature, $r_\ell(t)=r_\ell$, we recover well-known expressions for a time-independent setup with the rescaled temperature $r_\ell T_\ell$. Moreover, we can now appreciate the importance of including the scaling factor $r_\ell(t) s_\ell(t,\omega)$ in the interaction Hamiltonian (\ref{eq:intH}). Without this term, the spectral function of the baths and thus the self-energy, would not be rescaled properly, resulting in an effectively time-dependent spectral density. We recall again that we only modulate the temperature in the left bath.

Finally, with the spectral function above, we can evaluate the free Green's functions of the bath, which become
\begin{equation}
\mathcal{D}^>(\tau)=-2 i\alpha \left[\beta ^{-3}\zeta\!\left(3,1\!+\!\frac{\omega _c^{-1}\!+\!i
\tau}{\beta}\right)\!+\!\left(\omega _c^{-1}\!+\!i\tau\right)^{-3}\right],
\end{equation}
and
\begin{equation}
\mathcal{D}^R(\tau)=-2 i\alpha\theta(\tau) \left(\omega _c^{-1}+i
\tau\right)^{-3},
\end{equation}
where $\zeta(s,x)=\sum_{n=0}^\infty 1/(n+x)^s$ is the Hurwitz zeta function, and we have introduced the time difference $\tau=t-t'$. We then use the Langreth rules~\cite{Haug2008} to obtain the Dyson equations for the various Keldysh components of the complex contour Dyson equation in Eq.~(\ref{eq:MFDyson}), which we solve numerically as in Refs.~\cite{Talarico2019,Talarico2020}.

\section{DC heat transport}
\label{sec:DC}
A single qubit coupled to two heat baths at different temperatures has been predicted to show non-reciprocal transfer of energy. This feature has recently been exploited to engineer a heat-valve that can control the flow of heat~\cite{Ronzani2018,Senior2020}. In analogy with electrical circuits, one may consider such a device as a thermal diode, which could form a basic building block of thermal circuits, where temperature gradients and heat currents play the roles of voltage biases and electrical currents. One common figure-of-merit to quantify non-reciprocal transport is the difference between the forward and backward heat currents, defined as the response to a positive or negative temperature difference across the device with the average temperature kept constant. In the next section, we keep the right bath at the ground temperature $T_g$, while the left one is driven periodically with amplitude $\Delta T$ around~$T_g$. Consequently, we here define the forward and backward heat currents as the ones that are produced with the left bath held at the constant temperatures $T_g+\Delta T$ and $T_g-\Delta T$, respectively, while the right one has the fixed temperature $T_g$. (This definition is different from the most common approach of switching the temperatures of the baths, while keeping their average temperature constant.) For linear electrical circuits, the two cases would be equivalent, since a ground voltage would not affect the transport. By contrast, in our case,  non-reciprocal features arise due to the non-linear response of the qubit to the temperature bias.

As we argue now, the transport is always reciprocal in linear response regardless of what definition we use.
To see this, we find the stationary heat current from Eq.~(\ref{eq:hcurdef}) at long times, assuming that a stationary state has been reached and taking into account the conservation of energy, such that $J_L^{S}+J_R^{S}=0$ for the stationary currents based on our sign conventions. We then find
\begin{equation}
J_R^{S}=\frac{\alpha_L \alpha_R}{4\pi(\alpha_L +\alpha_R)}\int\limits_0^\infty \mathrm{d}\omega \omega I (\omega) \chi(\omega) [n_R(\omega)-n_L(\omega)],
\label{eq:nesscurr}
    \end{equation}
where 
\begin{equation}
\chi(\omega)=K_z^>(\omega)-K_z^<(\omega)
\end{equation}
is the susceptibility of the qubit in frequency space, and the Bose-Einstein distribution of each bath is denoted as
$n_\ell(\omega)$. Both bath spectral functions are given by Eq.~(\ref{eq:bathspectra}) with different coupling strengths, $\alpha_{L/R}$, so that $I(\omega)$ above is given by Eq.~(\ref{eq:bathspectra}) with $\alpha=1$ and the coupling strengths instead enter in front of the integral. From Eq.~(\ref{eq:nesscurr}), one can show that the forward and backward currents coincide in linear response, regardless of the definition of the temperature bias. Here, linear response is defined as having a small temperature bias compared to the ground temperature, $\Delta T\ll T_g$.  This finding implies that the system does not behave like a thermal diode in the linear-response regime, and we have to apply larger temperature biases to observe rectification. 

In the following, we consider the non-linear regime, where the definition of the temperature bias becomes important. In particular, we will explore two phenomena with polarity inversion on the diode, where the direction of the rectification changes as a function of a control parameter. In one case, we observe polarity inversion for weak couplings as the base temperature $T_g$ is lowered below the qubit energy $\varepsilon$. The other case occurs for high temperatures as the coupling strength is gradually increased. The first case is simply a consequence of how we define the bias, while the second one also occurs for the usual definition. We also note that if one simply switches the temperatures, there is no rectification for $\alpha_L=\alpha_R$, while with our definition, there may still be rectification, since our definition inherently is asymmetric. Here, we consider the symmetric case $\alpha_L=\alpha_R=\alpha$, and note that different couplings can either increase or decrease the difference between the forward and backward currents.

\begin{figure}
	\centering
	\includegraphics[width=0.95\columnwidth]{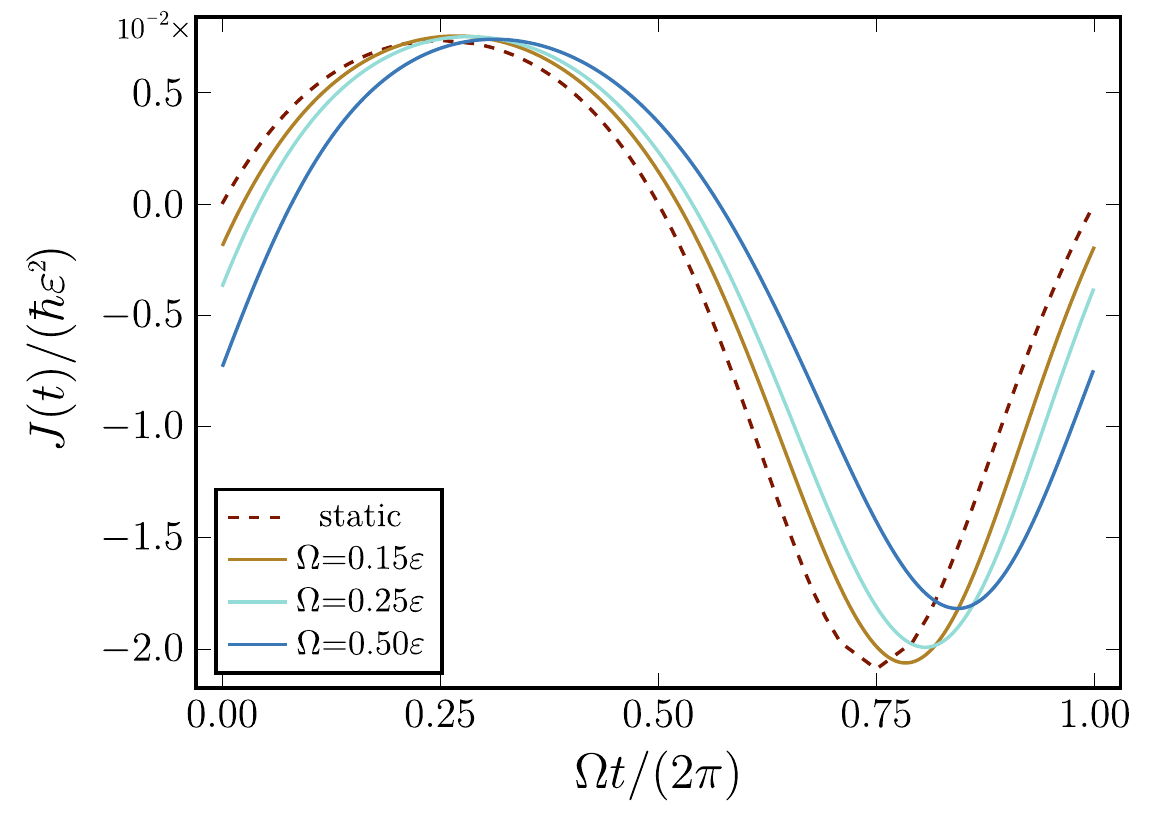}
	\caption{Time-dependent heat current. The dashed lines corresponds to the quasi-static limit, where the driving is so slow that the heat current at any time is given by Eq.~(\ref{eq:nesscurr}). The solid lines  correspond  to  different  driving  frequencies. The ground temperature and the driving amplitude are $k_B T_g=2.5\hbar \varepsilon$ and $\Delta T=0.6T_g$, respectively, and $\alpha=0.05$.}	\label{fig:threctdrive}
\end{figure}

In Fig.~\ref{fig:diode}a, we show the difference between the forward and backward currents as a function of the coupling strength and the ground temperature. As a function of the coupling strength, the polarity of the diode inverts close to the Toulouse point at $\alpha=1/2$. To understand this behavior, we note that the forward and backward susceptibilities in Fig.~\ref{fig:diode}b coincide exactly at the Toulouse point. However, the polarity changes slightly before the Toulouse point, because the heat current is given not only by the susceptibility but also by the difference of the Bose-Einstein distributions. This behaviour would also occur for the usual definition of the temperature bias. By contrast, the polarity inversion that can be seen as a function of temperature can be related to our specific definition of the bias and would not normally occur.

The microscopic origin of the polarity inversion can be understood by considering the imaginary part of the susceptibility in Fig.~\ref{fig:diode}b for different coupling strengths. The imaginary part of the susceptibility quantifies the response of the qubit to external forces at a given frequency~\cite{liu2016green}, and with a small susceptibility the qubit interacts only weakly with the environments, resulting in a small energy transfer. We note that the susceptibility can be related to a scattering matrix~\cite{belyansky2020transport}, and to the structure factor through fluctuation-dissipation theorems~\cite{weiss2012quantum}, which in turn quantifies the scattering of photons due to the interaction with the qubit \cite{Dattagupta_1989}.

\begin{figure}
\centering
\includegraphics[width=0.95\columnwidth]{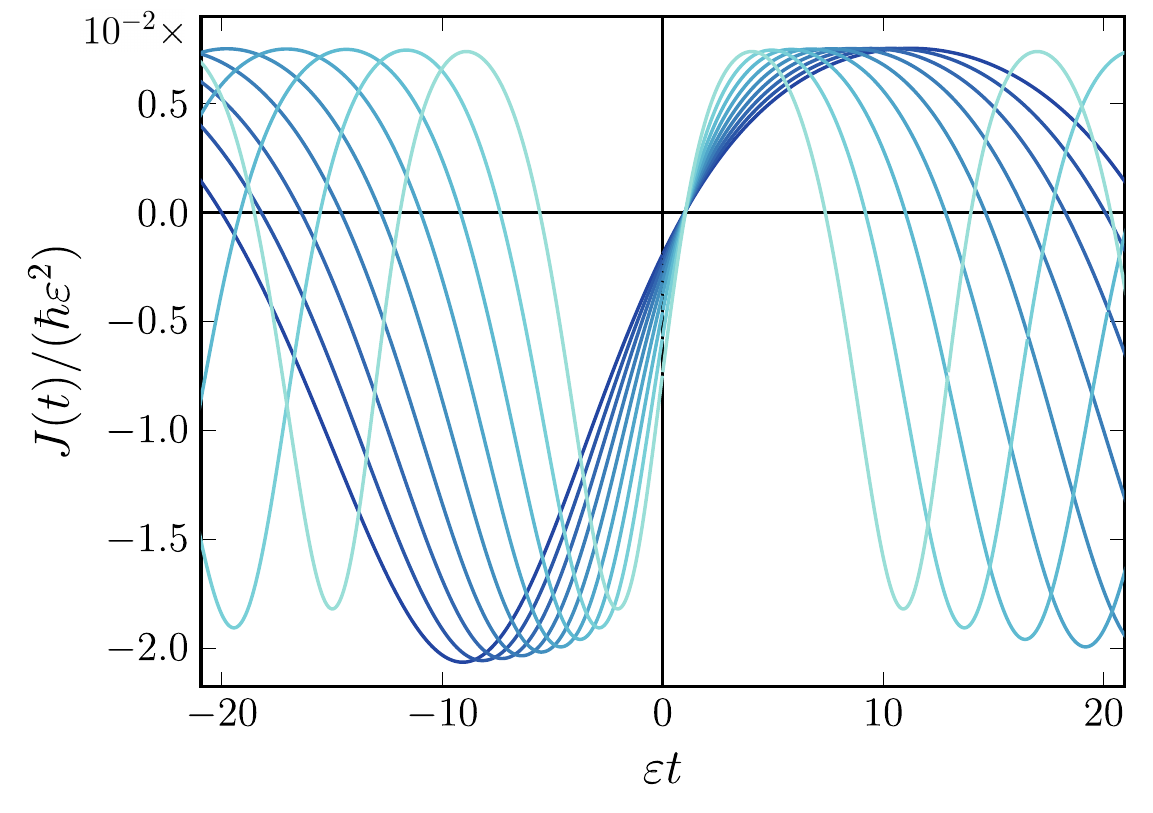}
\caption{Heat currents and phase shifts. We show time-dependent heat currents with driving frequencies increasing from $\Omega = 0.15\varepsilon$ (dark blue) to $\Omega = 0.5\varepsilon$ (light blue) in equidistant steps for the period. The other parameters are the same as in Fig.~\ref{fig:threctdrive}.}\label{fig:phase}
\end{figure}

\section{AC heat transport}
\label{sec:AC}

We now turn to the situation, where the qubit is driven by a periodically varying temperature of the left bath,
\begin{equation}
T(t)=T_g+\Delta T\sin(\Omega t),
\end{equation}
where $T_g$ is the base temperature of the reservoirs, and $\Delta T$ is the amplitude of the oscillations with frequency~$\Omega$. In what follows, we consider the qubit as a two-terminal element that processes an input signal, given by the periodic temperature modulation, and returns an output signal in terms of the time-dependent heat current in the right reservoir. In conventional electronics, a diode in series with a resistor functions as a half-wave rectifier for periodic voltage modulations. We now analyze the thermally driven qubit from a similar perspective and develop an equivalent circuit. We focus on high temperatures as the low-temperature regime displays very different features, which are beyond the scope of this work.

Figure~\ref{fig:threctdrive} shows the time-dependent current over a period of the drive for different driving frequencies. We also show the current obtained from Eq.~(\ref{eq:nesscurr}) using the instantaneous temperature corresponding to the quasi-static limit of low driving frequencies. First, we note that a phase shift between the input and output signals develops with increasing driving frequency. Thus, when driven by a periodic input signal, the circuit does not just behave as a thermal diode, but it also exhibits retardation effects. We also see that the rectification of the signal decreases with increasing driving frequency, leading to reduced maximally negative currents. We will come back to the characterization of the rectification, but we start by exploring the phase shift in more detail. 

In Fig.~\ref{fig:phase}, we again show the time-dependent heat current for different driving frequencies, however, without rescaling the time by the driving frequency. We can then observe that all signals cross at the same time after $t=0$, showing that the phase shift is proportional to the driving frequency. Such a phase shift is reminiscent of an $RC$-time in an electronic circuit, and we thus suggest the equivalent circuit in Fig.~\ref{fig:scheme}b with a resistor in parallel with a capacitor. For that circuit, the impedance reads $Z(\omega)=Re^{-i\phi(\omega)}$, where $\phi(\omega)=\arctan(\omega RC)$ is  the frequency-dependent phase shift of the output signal. At low frequencies, we then have $\phi(\omega)\simeq \omega RC$, which is consistent with our observation of a phase shift that is proportional to the driving frequency~$\Omega$. (We note that the resistor cannot be in series with the capacitor, since there would be no current with a constant bias.) 

\begin{figure}
    \centering
    \includegraphics[width=0.95\columnwidth]{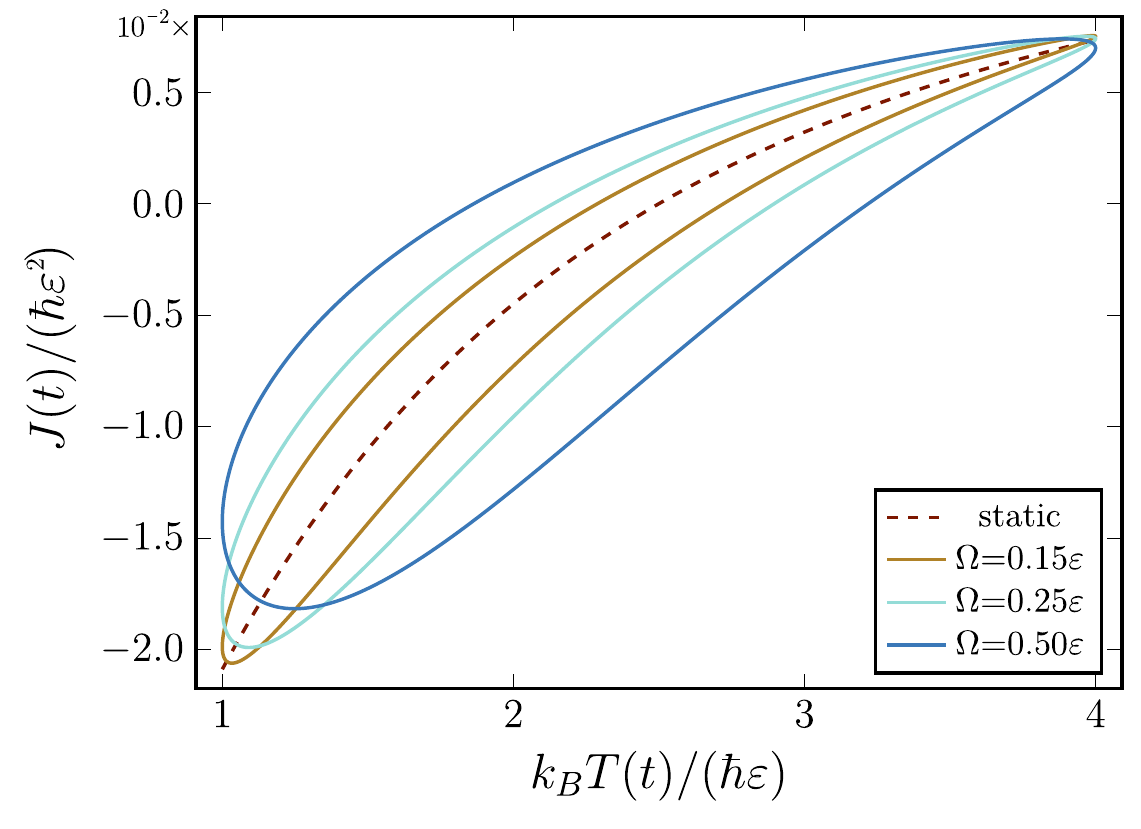}
\caption{Parametric plot of temperature and heat current. The dashed line shows results for the quasi-static limit, where the heat current is in sync with the temperature bias. The solid lines correspond to different driving frequencies. The base temperature is $k_B T_g=2.5\hbar \varepsilon$, the amplitude of the driving is $\Delta T=0.6T_g$, and the coupling strength is $\alpha=0.05$.}\label{fig:TJpar}
\end{figure}

To better illustrate the phase shift, we show in Fig.~\ref{fig:TJpar} a parametric plot of the temperature and the heat current. In the quasi-static limit, the input and output signals are in sync, and the area enclosed by the parametric curve vanishes. By contrast, with increasing driving frequency, a finite area builds up due to the phase shift between the two signals. In addition, the major axis of the ellipse-like shapes gradually rotates upwards, away from the quasi-static result. This behavior is consistent with the circuit in Fig.~\ref{fig:scheme}b. Eventually, at higher frequencies, we expect that a current can flow through the circuit in both directions via the capacitor despite the diode in the other branch. As a consequence, the rectification will be lost. 

\begin{figure}
    \centering
    \includegraphics[width=0.97\columnwidth]{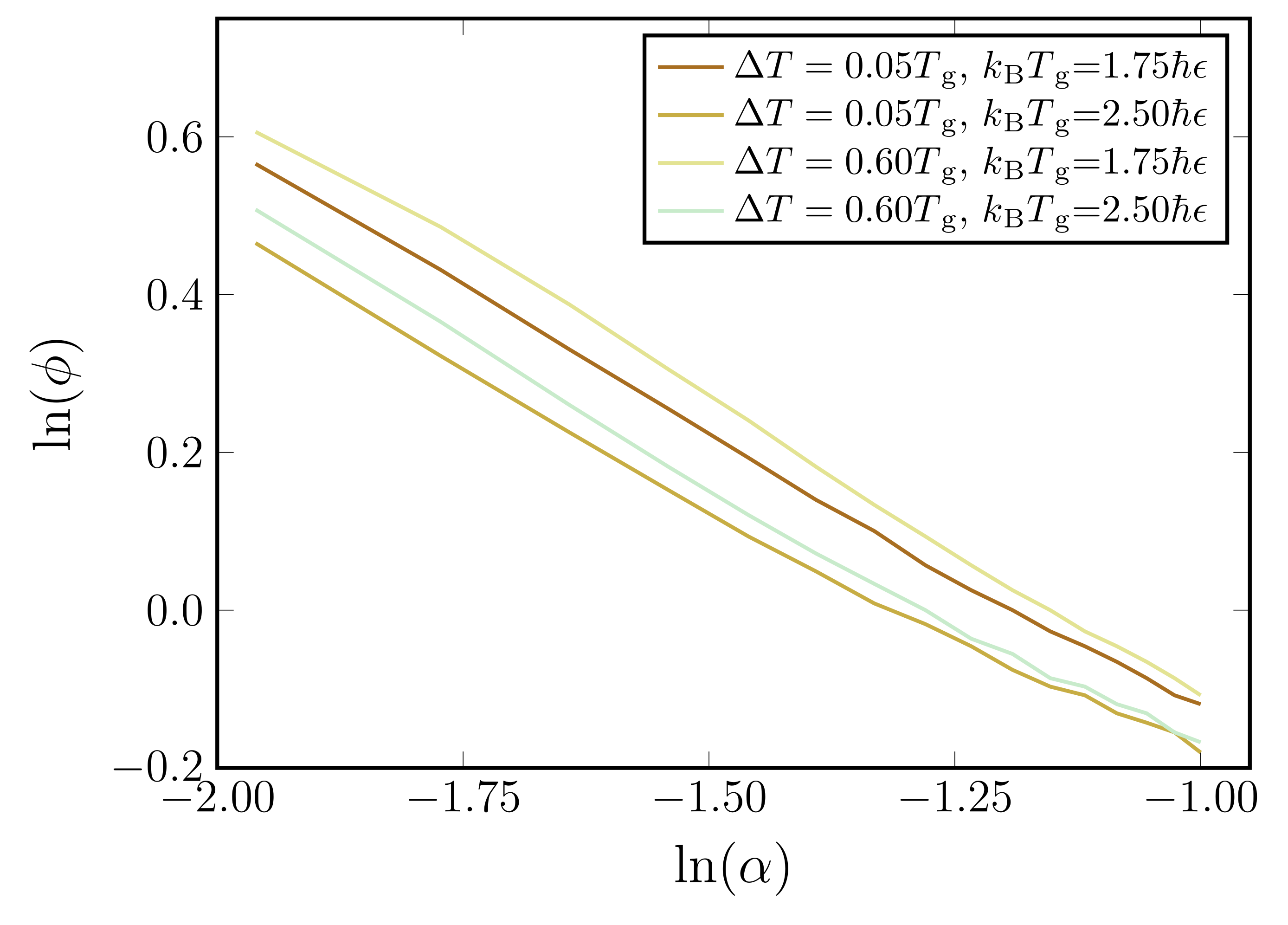}
    \caption{Input-output phase shift. The phase shift between the input and the output signal as a function of the coupling~$\alpha<0.5$ with the driving frequency $\Omega=0.2\varepsilon$. The phase shift approximately follows a power law. }
    \label{fig:phase_shift}
\end{figure}

Figure~\ref{fig:phase_shift} shows how the phase shift depends on the coupling strength, which also changes the resistance of the thermal circuit. In particular, for low couplings, $\alpha\sim 0.1$,  the current should increase linearly with increasing coupling, which thus lowers the resistance linearly. However, Figure~\ref{fig:phase_shift} shows that the phase shift to a good approximation displays a non-linear power law dependence on the coupling. This behavior shows that the capacitance also depends on the coupling strength following a power law.

\begin{figure}
	\centering
	\includegraphics[width=0.95\columnwidth]{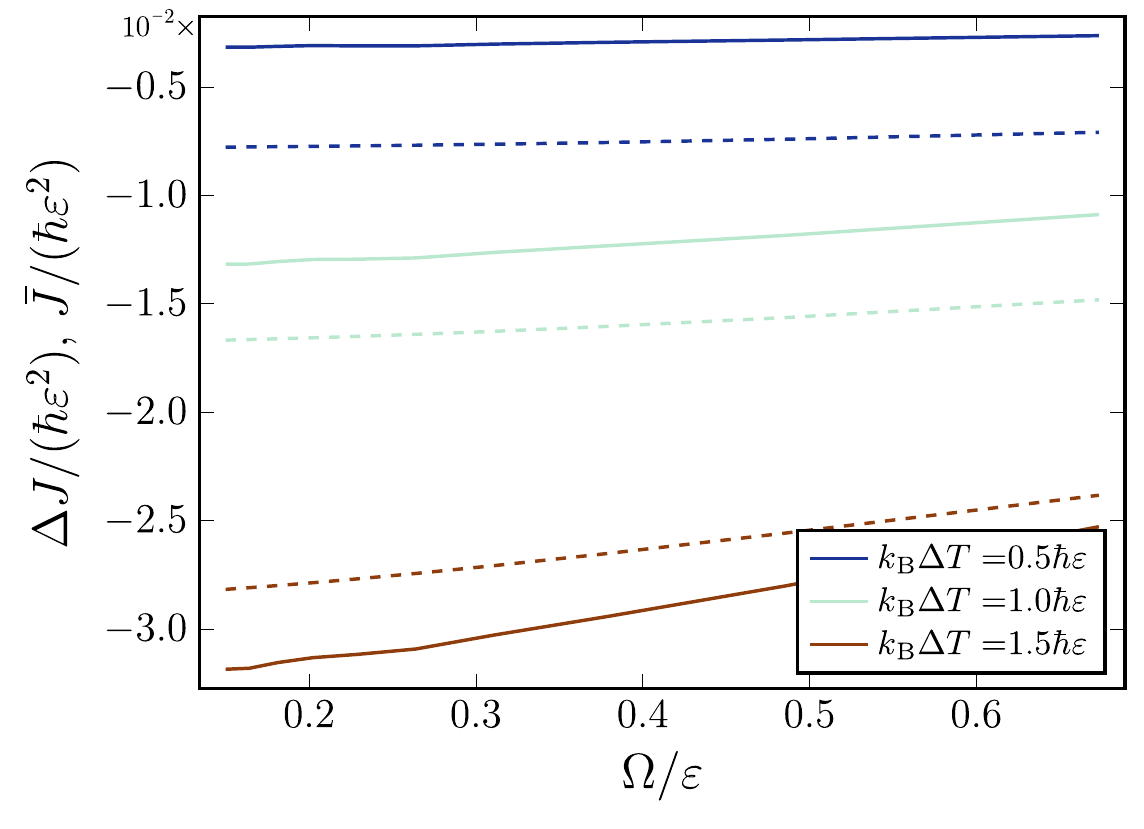}	
	\caption{Rectification of the input. Solid lines show the period-averaged heat current, while the dashed lines correspond to the difference between the minimum and maximum output. The ground temperature is $k_B T_g=2.5\hbar \varepsilon$, and $\alpha = 0.05$.}\label{fig:threctchar}
\end{figure}

In Fig.~\ref{fig:threctchar}, we return to the rectification in the system. The rectification can be characterized by two figures of merit: The period-averaged heat current, which we denote by $\bar{J}$, and the difference between the maximal negative and positive currents, $\Delta J$. These quantities may be used to estimate how well the output signal can be distinguished from the background noise, which is important in view of possible applications for digitizing the output signal for thermotronic applications. In Fig.~\ref{fig:threctchar}, we show the two quantities as functions of the driving frequency for three different amplitudes of the temperature oscillations, and we observe that the loss of rectification with increasing driving frequency is qualitatively similar for the three amplitudes and both figures of merit. 

Finally, we discuss possible parameter values for our calculations. Our results are based on a generic non-equilibrium spin-boson model, which potentially can be realized in many different physical platforms, for example using superconducting qubits \cite{Krantz2019}, atoms in optical lattices \cite{Brantut2013,Rossnagel2016}, electron spins in quantum dots \cite{Hanson2007}, or any other realization of a quantum two-level system coupled to external heat reservoirs. To be specific, we consider the spin of a trapped electron for which the tunable qubit spacing would be around $\varepsilon\simeq 500$ MHz in a magnetic field of 100 mT. This qubit splitting corresponds to a temperature of $T=\hbar\varepsilon/k_B\simeq 30$ mK, which certainly is reachable, more so, if we consider a temperature that is two or three times higher. \revision{In addition, the driving frequencies would be on the order of $100$-$200$ MHz, which certainly should be achievable with current technology. For comparison, voltages can be modulated at much higher frequencies of about 1-10 GHz, and a sample may be heated up by simply running a current through a resistor. Since electron and phonon relaxation times are typically much faster, the heat reservoirs will quickly equilibrate and follow the desired temperature modulations.  } 

\section{Conclusions}\label{sec:concl}
We have theoretically investigated the thermal transport in a quantum two-level system driven by a time-dependent temperature difference. To this end, we have extended a non-equilibrium Green's functions approach for static setups to include a time-dependent temperature. Based on this methodology, we have characterized the dynamic thermal properties of the quantum two-level system and showed that it can operate as a thermal half-way rectifier. We have proposed an equivalent circuit model of the setup consisting of a thermal diode in parallel with a capacitor. The thermal properties of the device are related to the quantum nature of the system, which must be accounted for in the design of thermal circuits at the nano scale. The method presented here can be extended to setups with more than two heat baths, for example, a three-terminal setup such as a thermal transistor. It would be also be interesting to investigate several coupled two-level systems, which may provide a means to enhance the asymmetry and rectification of a device. In practice, the spin-boson setup that we have considered can be implemented in many different ways, and we thus end by passing the baton to the experimentalists to identify the best systems to realize these ideas.

\begin{acknowledgements}
We thank M.~Moskalets and D.~S\'anchez for useful discussions and acknowledge support from  Academy of Finland through the Finnish Centre of Excellence in Quantum Technology (project nos.~312057 and 312299) and project nos.~308515 and~318937. Numerical calculations were performed using the Finnish CSC facilities under the project ``Thermoelectric effects in nanoscale devices'' (project no.~2000962).
\end{acknowledgements}

\appendix

\section{Heat current}
\label{app:hcur}
Here we describe the calculations leading to the expression for the heat current in Eq.~(\ref{eq:hcurdef}). The average change of the energy in the right bath $R$ is defined as 
\begin{equation}
J_R(t)=-i\left\langle[\hat{H}(t),\hat{H}_R]\right\rangle.
\end{equation}
Now, using the expressions for the total and the free Hamiltonian of the bath, we obtain
\begin{equation}
J_R(t)=2{\rm Re}\left[\sum\limits_{k} \omega_{R k}(t)g_{R k} M_{R kz}^<(t,t)\right],
\label{eq:hcrdet}
\end{equation}
where we have defined the mixed Green's function 
\begin{equation}
M_{R kz}(z,z')=-i\left\langle \mathcal{T}_\gamma \hat{a}_{R k}(z)\hat{\sigma}_z(z')\right\rangle.    
\end{equation}
For non-interacting baths, it is possible to derive an expression for the mixed Green's function in terms of the free bath propagator $\mathcal{D}(z,z')$ and the interacting Green's function of the qubit $K(z,z')$.
This procedure closely follows Refs.~\cite{liu2017,velizhanin2010meir} and yields
\begin{equation}
M_{R kz}(z,z')=g_{Rk}\int d\bar{z}\;\mathcal{D}(z,\bar{z})K_z(\bar{z},z').
\label{eq:mixed}
\end{equation}
Finally, by inserting this expression into the one for the heat current and projecting it onto the real-time axis, we arrive at Eq.~(\ref{eq:hcurdef}) for the heat current. 

\section{Evaluation of self-energies}
\label{app:dyson}

Here, we find the lesser and greater components of the self-energy and the correlations functions $B(t_1,t_2)$ for each bath. The lesser/greater self-energies read
\begin{equation}
\begin{split}
\Sigma^{>/<}(\tau,\tau')=\frac{i}{4}&\left[ G_{x0}^{>/<}(\tau,\tau') B_{x}^{>/<}(\tau,\tau')\right.\\
&\left.+G_{y0}^{>/<}(\tau,\tau') B_{y}^{>/<}(\tau,\tau')\right],
\end{split}
\label{eq:MFSE}
\end{equation}
where the qubit's Green's functions are just the imaginary unit $i$, since the Hamiltonian of the qubit in the polaron frame vanishes, and we have introduced the additional bath correlation functions
\begin{equation}
\begin{split}
B_{x}(\tau,\tau')&=-i\left\langle \mathcal{T}_\gamma\cos\hat{\Omega}(\tau)\cos\hat{\Omega}(\tau')\right\rangle,\\
B_{y}(\tau,\tau')&=-i\left\langle \mathcal{T}_\gamma\sin\hat{\Omega}(\tau)\sin\hat{\Omega}(\tau')\right\rangle,
\end{split}
\end{equation}
where $\hat{\Omega}$ is defined in Eq.~(\ref{eq:Omega}). The greater and lesser components of the free bath Green's functions in the polaron frame are given by
\begin{equation}
B^>(t,t^\prime)=-i\begin{pmatrix}
\langle \cos{\hat \Omega(t)}\cos{\hat\Omega(t^\prime)}\rangle & \langle \cos{\hat\Omega(t)}\sin{\hat\Omega(t^\prime)}\rangle \\
\langle \sin{\hat\Omega(t)}\cos{\hat\Omega(t^\prime)}\rangle  & \langle \sin{\hat\Omega(t)}\sin{\hat\Omega(t^\prime)}\rangle  
\end{pmatrix}
\end{equation}
with $B^>(t,t^\prime)=[B^{<}(t,t^\prime)]^*$. 

Next, we need to evaluate the four types of Green's functions involving combinations of trigonometric functions of $\hat\Omega$. These correlation functions can be expressed in terms of the functions 
\begin{equation}
\Phi_{mn}(t,t')=\langle e^{m i \hat \Omega(t)}e^{n i \hat\Omega(t')}\rangle,
\end{equation}
where $n$ and $m$ are integers. In the Heisenberg picture, we can write the annihilation operators of the bath as
\begin{equation}
\hat a_{\ell k}(t)=\hat a_{\ell k} e^{-i \omega_{\ell k} R_\ell(t)},
\end{equation}
where we have defined the time-integrated rescaling of the inverse temperature as
\begin{equation}
R_\ell(t)\equiv \int\limits_0^t \mathrm{d} s  r_\ell(s).
\end{equation}
We then find the expression
\begin{equation}
\begin{split}
\Phi_{n,m}(t,t')&=\prod_{k,\ell}(1-e^{\beta \omega_{\ell k}})\sum\limits_{n_k=0}^\infty e^{-\beta n_k \omega_{\ell k}}\\
&\times\left\langle{n_k}\right|e^{n\nu(t,\omega_{\ell k}) \hat p_{\ell k}(t)}e^{m\nu(t', \omega_{\ell k}) \hat p_{\ell k}(t')}\left|{n_k}\right\rangle
\end{split}
\label{eq:Phi_nm}
\end{equation} 
where $\beta$ is inverse temperature (before the rescaling), the eigenstates of the quantum harmonic oscillator are denoted as $\left|{n_k}\right\rangle$, and we have defined 
\begin{equation}
 \nu(t,\omega_{\ell k})=-2  \frac{g_{\ell k}}{\omega_{\ell k}} s(t,\omega_{\ell k})
 \label{eq:nu}
\end{equation} 
together with the operators 
\begin{equation}
\hat p_{\ell k}(t)=\hat a_{\ell k}^{\dagger} e^{i R_\ell(t) \omega_{\ell k} }-\hat a_{\ell k} e^{-i R_\ell(t) \omega_{\ell k}}.
\end{equation} 

Next, we factor out the time-dependence and normal-order the operators. To this end, we recall that
\begin{equation}
e^{\hat A+\hat B}=e^{\hat A} e^{\hat B} e^{-[\hat A,\hat B]/2}
\end{equation}
for pairs of operators, whose nested commutators vanish, $[[\hat A,\hat B],\hat A]=[[\hat A,\hat B],\hat B]=0$. Considering the product of exponentials in Eq.~(\ref{eq:Phi_nm}), we then find
\begin{equation}
e^{n\nu_1 \hat p_1}e^{m\nu_2 \hat  p_2}= e^{n\nu_1 \hat  p_1+m\nu_2 \hat  p_2} e^{nm\nu_1 \nu_2 [\hat p_1,\hat p_2]/2},
\end{equation}
having used subscripts to represent the different variables of the functions. We can rewrite these expressions as
\begin{equation}
 n\nu_1 \hat p_1+m\nu_2 \hat p_2= \xi \hat a^\dagger-\xi^* \hat a, 
\end{equation}
and
\begin{equation}
\begin{split}
[\hat p_1,\hat p_2]&=e^{-i\tilde R_\ell(t,t') \omega }-e^{i\tilde R_\ell(t,t') \omega }\\
&=- 2i\sin{(\tilde R_\ell(t,t') \omega)},
\end{split}
\end{equation}
having defined  $\xi=n\nu_1 e^{iR_\ell(t)\omega }+m\nu_2 e^{iR_\ell(t')\omega}$ and 
\begin{equation}
\tilde R_\ell(t,t')=R_\ell(t)-R_\ell(t')=\int\limits_{t'}^{t} \mathrm{d} s r_\ell(s).
\end{equation}
We then find
\begin{equation}
e^{n\nu_1 \hat p_1}e^{n\nu_2 \hat p_2}=e^{-|\xi|^2/2} e^{\xi \hat a^\dagger}e^{-\xi^* \hat a}e^{-i nm\nu_1 \nu_2 \sin(\tilde R_\ell \omega )},
\end{equation}
and by inserting this expression into Eq.~(\ref{eq:Phi_nm}), we obtain
\begin{widetext}
\begin{equation}
\begin{split}
\Phi_{n,m}(t,t')&=\prod_{k,\ell} e^{-i nm\nu_1^k \nu_2^k  \sin(\tilde R_\ell\omega_{\ell k} )}e^{|\xi|^2} (1-e^{\beta \omega_{\ell k}})\sum_{n_k=0}^\infty  e^{-\beta n_k \omega_{\ell k}}\left\langle n_k\right| e^{\xi \hat a^\dagger}e^{-\xi^* \hat a}\left| n_k\right\rangle\\
&=\prod_{k,\ell} e^{-i nm\nu_1^k \nu_2^k \sin(\tilde R_\ell\omega_{\ell k} )}e^{-|\xi|^2/2}e^{-|\xi|^2 n_B(\omega_{\ell k})}\equiv e^{K_{n,m}(t,t')},
\end{split}
\end{equation}
where we have introduced the matrix elements
\begin{equation}
\begin{split}
K_{n,m}(t,t')&=\sum_{k,\ell} -i nm \nu_1^k \nu_2^k \sin(\tilde{R}_\ell \omega_{\ell k} )-\frac{1}{2}|\xi|^2(1+2 n_B(\omega_{\ell k}))\\
&=\sum_{k,\ell} 4\frac{g_{\ell k}^2}{\omega_{\ell k}^2}\left[-i n m s(t)s(t')\sin(\tilde R_{\ell} \omega_{\ell k} )+\left(\frac{s^2(t)+s^2(t')}{2}+n m s(t_1)s(t_2)\cos(\tilde{R}_\ell \omega_{\ell k}  )\right)\coth(\beta\omega_{\ell k} /2) \right],
\end{split}
\end{equation}
having used Eq.~(\ref{eq:nu}) in the second line. This expression can be written as a sum over the baths, $K_{n,m}(t,t')=\sum_{\ell}K^{(\ell)}_{n,m}(t,t')$, and we focus now on the contribution from one of the baths. Assuming a continuous density of states, we may replace the sum by an integral and write
\begin{equation}
K^{(\ell)}_{n,m}(t,t')=\int \limits_0^\infty\mathrm{d}\omega \frac{2}{\pi}\frac{I_{\ell}(\omega)}{\omega^2}\left[-i n m s(t)s(t')\sin(\tilde R_{\ell} \omega )+\left(\frac{s^2(t)+s^2(t')}{2}+n m s(t_1)s(t_2)\cos(\tilde R_{\ell} \omega )\right)\coth(\beta\omega/2) \right],
\end{equation}
where $I_{\ell}(\omega)=2\pi \sum_k g_{\ell k}^2 \delta(\omega-\omega_{\ell k})$ is the spectral density of the bath, which we take to be of the form
\begin{equation}
I_{\ell}(\omega)=\alpha_{\ell} \pi  \omega^p \omega_c^{1-p} e^{-\omega/\omega_c}
\end{equation}
as in Eq.~(\ref{eq:bathspectra}). We now see that the rescaled spectral density can be written as
\begin{equation}
\frac{I_{\ell}(r_{\ell}\omega)}{s^2_{\ell}r^2_{\ell}}=\frac{r_{\ell}^{p-2}e^{(1-r_{\ell})\omega/\omega_c}}{s^2_{\ell}} I_{\ell}(\omega),
\end{equation}
and requiring that it must equal the original spectral density $I_{\ell}(\omega)$, we set the fraction on the right hand side equal to unity, which in turn fixes the dependence of $s_{\ell}(t,\omega)$ on $r_{\ell}(t)$ as
\begin{equation}
s_{\ell}(t,\omega)=[r_{\ell}(t)]^{\frac{p-2}{2}}e^{ \frac{1-r_{\ell}(t)}{2}\omega/\omega_c}.
\end{equation}

As an important sanity check, we find in the case, where $r_{\ell}(t)=r_{\ell}$ is constant, the expression
\begin{equation}
\begin{split}
K^{(\ell)}_{n,m}(\tau)&=2\alpha_{\ell}\int \limits_0^\infty (r_{\ell}\mathrm{d}\omega)  (r_{\ell}\omega)^{(p-2)} \omega_c^{1-p} e^{-r_{\ell}\omega/\omega_c}\left[-i n m  \sin(r_{\ell}\omega \tau)+\left[1+n m \cos(r_{\ell}\omega \tau)\right]\coth(r_{\ell}\omega (\beta/r_{\ell})/2) \right]\\
&=2\alpha_{\ell}\int \limits_0^\infty \mathrm{d}\omega  \omega^{(p-2)} \omega_c^{1-p} e^{-\omega/\omega_c}\left[-i n m  \sin(\omega \tau)+\left[1+n m \cos(\omega \tau)\right]\coth(\omega (\beta/r_{\ell})/2) \right],
\end{split}
\end{equation}
which is a function of the time difference $\tau=t-t'$ only and exactly corresponds to a bath with the constant rescaled inverse temperature $\beta/r_{\ell}$. 

The final time-dependent expression reads 
\begin{equation}
\begin{split}
&K^{(\ell)}_{n,m}(t,t')= 2\alpha_{\ell} \int \limits_0^\infty \mathrm{d}\omega \omega^{(p-2)} \omega_c^{1-p}\bigg[-i n m \sqrt{[r_{\ell}(t)r_{\ell}(t')]^{p-1}}e^{-\left(\frac{r_{\ell}(t)+r_{\ell}(t')}{2}\right)\frac{\omega}{\omega_c}}\sin( \tilde R_{\ell}\omega)\\
&+\left(\frac{[r_{\ell}(t)]^{p-1}e^{-r_{\ell}(t)\frac{\omega}{\omega_c}}+[r_{\ell}(t')]^{p-1}e^{-r_{\ell}(t')\frac{\omega}{\omega_c}}}{2}+n m \sqrt{[r_{\ell}(t)r_{\ell}(t')]^{p-1}}e^{-\left(\frac{r_{\ell}(t)+r_{\ell}(t')}{2}\right)\frac{\omega}{\omega_c}} \cos( \tilde R_{\ell}\omega)\right)\coth(\beta \omega /2) \bigg]
\end{split}
\end{equation}
To evaluate the Green's functions in the main text, we only need the terms $K^{(\ell)}_{1,-1}$ and $K^{(\ell)}_{-1,1}$ with the properties $K^{(\ell)}_{1,-1}=K^{(\ell)}_{-1,1}\equiv K^{(\ell)}$, such that only $K^{(\ell)}(t,t')$ needs to be evaluated. Using
\begin{equation}
\zeta (z,q)=\frac{1}{\Gamma(z)}\int\limits_0^\infty \mathrm d \omega\frac{\omega^{z-1}e^{-q \omega}}{1-e^{-\omega}},
\end{equation}
we can evaluate $K^{(\ell)}(t,t')$  analytically, and in the ohmic limit $p\rightarrow 1$ we find
\begin{equation}
K^{(\ell)}(t,t')= \alpha_{\ell}\ln\left(\frac{\beta ^2 \omega _c^2 \Gamma^2 [\frac{r_{\ell}(t)}{\beta  \omega _c}+1] \Gamma^2
   [\frac{r_{\ell}(t')}{\beta  \omega_c}+1]}{r_{\ell}(t) r_{\ell}(t') \Gamma^2 [c^{(\ell)}_+(t,t')] \Gamma^2[c^{(\ell)}_-(t,t')+1]}\right),
\end{equation}
where $\Gamma[x]$ is the gamma function, and the function $c^{(\ell)}_\pm(t,t')$ is defined in Eq.~(\ref{eq:c-func}) for a single heat bath. These expressions finally lead us to Eq.~(\ref{eq:Phi-func}), which allows us to calculate the heat current for a time-dependent temperature.
\end{widetext}


\begin{thebibliography}{50}%
	\makeatletter
	\providecommand \@ifxundefined [1]{%
		\@ifx{#1\undefined}
	}%
	\providecommand \@ifnum [1]{%
		\ifnum #1\expandafter \@firstoftwo
		\else \expandafter \@secondoftwo
		\fi
	}%
	\providecommand \@ifx [1]{%
		\ifx #1\expandafter \@firstoftwo
		\else \expandafter \@secondoftwo
		\fi
	}%
	\providecommand \natexlab [1]{#1}%
	\providecommand \enquote  [1]{``#1''}%
	\providecommand \bibnamefont  [1]{#1}%
	\providecommand \bibfnamefont [1]{#1}%
	\providecommand \citenamefont [1]{#1}%
	\providecommand \href@noop [0]{\@secondoftwo}%
	\providecommand \href [0]{\begingroup \@sanitize@url \@href}%
	\providecommand \@href[1]{\@@startlink{#1}\@@href}%
	\providecommand \@@href[1]{\endgroup#1\@@endlink}%
	\providecommand \@sanitize@url [0]{\catcode `\\12\catcode `\$12\catcode
		`\&12\catcode `\#12\catcode `\^12\catcode `\_12\catcode `\%12\relax}%
	\providecommand \@@startlink[1]{}%
	\providecommand \@@endlink[0]{}%
	\providecommand \url  [0]{\begingroup\@sanitize@url \@url }%
	\providecommand \@url [1]{\endgroup\@href {#1}{\urlprefix }}%
	\providecommand \urlprefix  [0]{URL }%
	\providecommand \Eprint [0]{\href }%
	\providecommand \doibase [0]{https://doi.org/}%
	\providecommand \selectlanguage [0]{\@gobble}%
	\providecommand \bibinfo  [0]{\@secondoftwo}%
	\providecommand \bibfield  [0]{\@secondoftwo}%
	\providecommand \translation [1]{[#1]}%
	\providecommand \BibitemOpen [0]{}%
	\providecommand \bibitemStop [0]{}%
	\providecommand \bibitemNoStop [0]{.\EOS\space}%
	\providecommand \EOS [0]{\spacefactor3000\relax}%
	\providecommand \BibitemShut  [1]{\csname bibitem#1\endcsname}%
	\let\auto@bib@innerbib\@empty
	\bibitem [{\citenamefont {Streetman}\ and\ \citenamefont
		{Banerjee}(2016)}]{Streetman2016}%
	\BibitemOpen
	\bibfield  {author} {\bibinfo {author} {\bibfnamefont {B.~G.}\ \bibnamefont
			{Streetman}}\ and\ \bibinfo {author} {\bibfnamefont {S.~K.}\ \bibnamefont
			{Banerjee}},\ }\href@noop {} {\emph {\bibinfo {title} {Solid State Electronic
				Devices}}}\ (\bibinfo  {publisher} {Pearson},\ \bibinfo {year}
	{2016})\BibitemShut {NoStop}%
	\bibitem [{\citenamefont {Giazotto}\ \emph {et~al.}(2006)\citenamefont
		{Giazotto}, \citenamefont {Heikkil\"a}, \citenamefont {Luukanen},
		\citenamefont {Savin},\ and\ \citenamefont {Pekola}}]{Giazotto2006}%
	\BibitemOpen
	\bibfield  {author} {\bibinfo {author} {\bibfnamefont {F.}~\bibnamefont
			{Giazotto}}, \bibinfo {author} {\bibfnamefont {T.~T.}\ \bibnamefont
			{Heikkil\"a}}, \bibinfo {author} {\bibfnamefont {A.}~\bibnamefont
			{Luukanen}}, \bibinfo {author} {\bibfnamefont {A.~M.}\ \bibnamefont
			{Savin}},\ and\ \bibinfo {author} {\bibfnamefont {J.~P.}\ \bibnamefont
			{Pekola}},\ }\bibfield  {title} {\bibinfo {title} {Opportunities for
			mesoscopics in thermometry and refrigeration: Physics and applications},\
	}\href {https://doi.org/10.1103/RevModPhys.78.217} {\bibfield  {journal}
		{\bibinfo  {journal} {Rev. Mod. Phys.}\ }\textbf {\bibinfo {volume} {78}},\
		\bibinfo {pages} {217} (\bibinfo {year} {2006})}\BibitemShut {NoStop}%
	\bibitem [{\citenamefont {Pekola}(2015)}]{Pekola2015}%
	\BibitemOpen
	\bibfield  {author} {\bibinfo {author} {\bibfnamefont {J.~P.}\ \bibnamefont
			{Pekola}},\ }\bibfield  {title} {\bibinfo {title} {Towards quantum
			thermodynamics in electronic circuits},\ }\href
	{https://doi.org/10.1038/nphys3169} {\bibfield  {journal} {\bibinfo
			{journal} {Nature Phys.}\ }\textbf {\bibinfo {volume} {11}},\ \bibinfo
		{pages} {118} (\bibinfo {year} {2015})}\BibitemShut {NoStop}%
	\bibitem [{\citenamefont {Jones}(2018)}]{Jones2018}%
	\BibitemOpen
	\bibfield  {author} {\bibinfo {author} {\bibfnamefont {N.}~\bibnamefont
			{Jones}},\ }\bibfield  {title} {\bibinfo {title} {How to stop data centres
			from gobbling up the world’s electricity},\ }\href
	{https://doi.org/10.1038/d41586-018-06610-y} {\bibfield  {journal} {\bibinfo
			{journal} {Nature}\ }\textbf {\bibinfo {volume} {561}},\ \bibinfo {pages}
		{163} (\bibinfo {year} {2018})}\BibitemShut {NoStop}%
	\bibitem [{\citenamefont {Shakouri}(2011)}]{Shakouri2011}%
	\BibitemOpen
	\bibfield  {author} {\bibinfo {author} {\bibfnamefont {A.}~\bibnamefont
			{Shakouri}},\ }\bibfield  {title} {\bibinfo {title} {{Recent Developments in
				Semiconductor Thermoelectric Physics and Materials}},\ }\href
	{https://doi.org/10.1146/annurev-matsci-062910-100445} {\bibfield  {journal}
		{\bibinfo  {journal} {Annu. Rev. Mater. Res.}\ }\textbf {\bibinfo {volume}
			{41}},\ \bibinfo {pages} {399} (\bibinfo {year} {2011})}\BibitemShut
	{NoStop}%
	\bibitem [{\citenamefont {Ben-Abdallah}\ and\ \citenamefont
		{Biehs}(2016)}]{Biehs2017}%
	\BibitemOpen
	\bibfield  {author} {\bibinfo {author} {\bibfnamefont {P.}~\bibnamefont
			{Ben-Abdallah}}\ and\ \bibinfo {author} {\bibfnamefont {S.-A.}\ \bibnamefont
			{Biehs}},\ }\bibfield  {title} {\bibinfo {title} {{Thermotronics: Towards
				Nanocircuits to Manage Radiative Heat Flux}},\ }\href
	{https://doi.org/https://doi.org/10.1515/zna-2016-0358} {\bibfield  {journal}
		{\bibinfo  {journal} {Z. Naturforsch. A}\ }\textbf {\bibinfo {volume} {72}},\
		\bibinfo {pages} {151} (\bibinfo {year} {2016})}\BibitemShut {NoStop}%
	\bibitem [{\citenamefont {Vining}(2009)}]{vining2009inconvenient}%
	\BibitemOpen
	\bibfield  {author} {\bibinfo {author} {\bibfnamefont {C.~B.}\ \bibnamefont
			{Vining}},\ }\bibfield  {title} {\bibinfo {title} {An inconvenient truth
			about thermoelectrics},\ }\href {https://doi.org/10.1038/nmat2361} {\bibfield
		{journal} {\bibinfo  {journal} {Nature Mat.}\ }\textbf {\bibinfo {volume}
			{8}},\ \bibinfo {pages} {83} (\bibinfo {year} {2009})}\BibitemShut {NoStop}%
	\bibitem [{\citenamefont {Dhar}(2008)}]{Dhar2008}%
	\BibitemOpen
	\bibfield  {author} {\bibinfo {author} {\bibfnamefont {A.}~\bibnamefont
			{Dhar}},\ }\bibfield  {title} {\bibinfo {title} {Heat transport in
			low-dimensional systems},\ }\href {https://doi.org/10.1080/00018730802538522}
	{\bibfield  {journal} {\bibinfo  {journal} {Adv. Phys.}\ }\textbf {\bibinfo
			{volume} {57}},\ \bibinfo {pages} {457} (\bibinfo {year} {2008})}\BibitemShut
	{NoStop}%
	\bibitem [{\citenamefont {Li}\ \emph {et~al.}(2012)\citenamefont {Li},
		\citenamefont {Ren}, \citenamefont {Wang}, \citenamefont {Zhang},
		\citenamefont {H\"anggi},\ and\ \citenamefont {Li}}]{hanggi2012}%
	\BibitemOpen
	\bibfield  {author} {\bibinfo {author} {\bibfnamefont {N.}~\bibnamefont
			{Li}}, \bibinfo {author} {\bibfnamefont {J.}~\bibnamefont {Ren}}, \bibinfo
		{author} {\bibfnamefont {L.}~\bibnamefont {Wang}}, \bibinfo {author}
		{\bibfnamefont {G.}~\bibnamefont {Zhang}}, \bibinfo {author} {\bibfnamefont
			{P.}~\bibnamefont {H\"anggi}},\ and\ \bibinfo {author} {\bibfnamefont
			{B.}~\bibnamefont {Li}},\ }\bibfield  {title} {\bibinfo {title} {Phononics:
			Manipulating heat flow with electronic analogs and beyond},\ }\href
	{https://doi.org/10.1103/RevModPhys.84.1045} {\bibfield  {journal} {\bibinfo
			{journal} {Rev. Mod. Phys.}\ }\textbf {\bibinfo {volume} {84}},\ \bibinfo
		{pages} {1045} (\bibinfo {year} {2012})}\BibitemShut {NoStop}%
	\bibitem [{\citenamefont {Wu}\ \emph {et~al.}(2009)\citenamefont {Wu},
		\citenamefont {Yu},\ and\ \citenamefont {Segal}}]{Wu2009}%
	\BibitemOpen
	\bibfield  {author} {\bibinfo {author} {\bibfnamefont {L.-A.}\ \bibnamefont
			{Wu}}, \bibinfo {author} {\bibfnamefont {C.~X.}\ \bibnamefont {Yu}},\ and\
		\bibinfo {author} {\bibfnamefont {D.}~\bibnamefont {Segal}},\ }\bibfield
	{title} {\bibinfo {title} {Nonlinear quantum heat transfer in hybrid
			structures: Sufficient conditions for thermal rectification},\ }\href
	{https://doi.org/10.1103/PhysRevE.80.041103} {\bibfield  {journal} {\bibinfo
			{journal} {Phys. Rev. E}\ }\textbf {\bibinfo {volume} {80}},\ \bibinfo
		{pages} {041103} (\bibinfo {year} {2009})}\BibitemShut {NoStop}%
	\bibitem [{\citenamefont {Pascal}\ \emph {et~al.}(2011)\citenamefont {Pascal},
		\citenamefont {Courtois},\ and\ \citenamefont {Hekking}}]{Pascal2011}%
	\BibitemOpen
	\bibfield  {author} {\bibinfo {author} {\bibfnamefont {L.~M.~A.}\
			\bibnamefont {Pascal}}, \bibinfo {author} {\bibfnamefont {H.}~\bibnamefont
			{Courtois}},\ and\ \bibinfo {author} {\bibfnamefont {F.~W.~J.}\ \bibnamefont
			{Hekking}},\ }\bibfield  {title} {\bibinfo {title} {Circuit approach to
			photonic heat transport},\ }\href
	{https://doi.org/10.1103/PhysRevB.83.125113} {\bibfield  {journal} {\bibinfo
			{journal} {Phys. Rev. B}\ }\textbf {\bibinfo {volume} {83}},\ \bibinfo
		{pages} {125113} (\bibinfo {year} {2011})}\BibitemShut {NoStop}%
	\bibitem [{\citenamefont {Ben-Abdallah}\ \emph {et~al.}(2011)\citenamefont
		{Ben-Abdallah}, \citenamefont {Biehs},\ and\ \citenamefont
		{Joulain}}]{Ben-Abdallah2011}%
	\BibitemOpen
	\bibfield  {author} {\bibinfo {author} {\bibfnamefont {P.}~\bibnamefont
			{Ben-Abdallah}}, \bibinfo {author} {\bibfnamefont {S.-A.}\ \bibnamefont
			{Biehs}},\ and\ \bibinfo {author} {\bibfnamefont {K.}~\bibnamefont
			{Joulain}},\ }\bibfield  {title} {\bibinfo {title} {{Many-Body Radiative Heat
				Transfer Theory}},\ }\href {https://doi.org/10.1103/PhysRevLett.107.114301}
	{\bibfield  {journal} {\bibinfo  {journal} {Phys. Rev. Lett.}\ }\textbf
		{\bibinfo {volume} {107}},\ \bibinfo {pages} {114301} (\bibinfo {year}
		{2011})}\BibitemShut {NoStop}%
	\bibitem [{\citenamefont {Brange}\ \emph {et~al.}(2019)\citenamefont {Brange},
		\citenamefont {Menczel},\ and\ \citenamefont {Flindt}}]{Brange2019}%
	\BibitemOpen
	\bibfield  {author} {\bibinfo {author} {\bibfnamefont {F.}~\bibnamefont
			{Brange}}, \bibinfo {author} {\bibfnamefont {P.}~\bibnamefont {Menczel}},\
		and\ \bibinfo {author} {\bibfnamefont {C.}~\bibnamefont {Flindt}},\
	}\bibfield  {title} {\bibinfo {title} {Photon counting statistics of a
			microwave cavity},\ }\href {https://doi.org/10.1103/PhysRevB.99.085418}
	{\bibfield  {journal} {\bibinfo  {journal} {Phys. Rev. B}\ }\textbf {\bibinfo
			{volume} {99}},\ \bibinfo {pages} {085418} (\bibinfo {year}
		{2019})}\BibitemShut {NoStop}%
	\bibitem [{\citenamefont {Ojanen}\ and\ \citenamefont
		{Jauho}(2008)}]{Ojanen2008}%
	\BibitemOpen
	\bibfield  {author} {\bibinfo {author} {\bibfnamefont {T.}~\bibnamefont
			{Ojanen}}\ and\ \bibinfo {author} {\bibfnamefont {A.-P.}\ \bibnamefont
			{Jauho}},\ }\bibfield  {title} {\bibinfo {title} {{Mesoscopic Photon Heat
				Transistor}},\ }\href {https://doi.org/10.1103/PhysRevLett.100.155902}
	{\bibfield  {journal} {\bibinfo  {journal} {Phys. Rev. Lett.}\ }\textbf
		{\bibinfo {volume} {100}},\ \bibinfo {pages} {155902} (\bibinfo {year}
		{2008})}\BibitemShut {NoStop}%
	\bibitem [{\citenamefont {Ruokola}\ \emph {et~al.}(2009)\citenamefont
		{Ruokola}, \citenamefont {Ojanen},\ and\ \citenamefont
		{Jauho}}]{Ruokola2009}%
	\BibitemOpen
	\bibfield  {author} {\bibinfo {author} {\bibfnamefont {T.}~\bibnamefont
			{Ruokola}}, \bibinfo {author} {\bibfnamefont {T.}~\bibnamefont {Ojanen}},\
		and\ \bibinfo {author} {\bibfnamefont {A.-P.}\ \bibnamefont {Jauho}},\
	}\bibfield  {title} {\bibinfo {title} {Thermal rectification in nonlinear
			quantum circuits},\ }\href {https://doi.org/10.1103/PhysRevB.79.144306}
	{\bibfield  {journal} {\bibinfo  {journal} {Phys. Rev. B}\ }\textbf {\bibinfo
			{volume} {79}},\ \bibinfo {pages} {144306} (\bibinfo {year}
		{2009})}\BibitemShut {NoStop}%
	\bibitem [{\citenamefont {Peotta}\ and\ \citenamefont
		{Di~Ventra}(2014)}]{Peotta2014}%
	\BibitemOpen
	\bibfield  {author} {\bibinfo {author} {\bibfnamefont {S.}~\bibnamefont
			{Peotta}}\ and\ \bibinfo {author} {\bibfnamefont {M.}~\bibnamefont
			{Di~Ventra}},\ }\bibfield  {title} {\bibinfo {title} {{Superconducting
				Memristors}},\ }\href {https://doi.org/10.1103/PhysRevApplied.2.034011}
	{\bibfield  {journal} {\bibinfo  {journal} {Phys. Rev. Applied}\ }\textbf
		{\bibinfo {volume} {2}},\ \bibinfo {pages} {034011} (\bibinfo {year}
		{2014})}\BibitemShut {NoStop}%
	\bibitem [{\citenamefont {Kubytskyi}\ \emph {et~al.}(2014)\citenamefont
		{Kubytskyi}, \citenamefont {Biehs},\ and\ \citenamefont
		{Ben-Abdallah}}]{Kubytsky2014}%
	\BibitemOpen
	\bibfield  {author} {\bibinfo {author} {\bibfnamefont {V.}~\bibnamefont
			{Kubytskyi}}, \bibinfo {author} {\bibfnamefont {S.-A.}\ \bibnamefont
			{Biehs}},\ and\ \bibinfo {author} {\bibfnamefont {P.}~\bibnamefont
			{Ben-Abdallah}},\ }\bibfield  {title} {\bibinfo {title} {{Radiative
				Bistability and Thermal Memory}},\ }\href
	{https://doi.org/10.1103/PhysRevLett.113.074301} {\bibfield  {journal}
		{\bibinfo  {journal} {Phys. Rev. Lett.}\ }\textbf {\bibinfo {volume} {113}},\
		\bibinfo {pages} {074301} (\bibinfo {year} {2014})}\BibitemShut {NoStop}%
	\bibitem [{\citenamefont {Ordonez-Miranda}\ \emph {et~al.}(2019)\citenamefont
		{Ordonez-Miranda}, \citenamefont {Ezzahri}, \citenamefont {Tiburcio-Moreno},
		\citenamefont {Joulain},\ and\ \citenamefont {Drevillon}}]{Ordonez2019}%
	\BibitemOpen
	\bibfield  {author} {\bibinfo {author} {\bibfnamefont {J.}~\bibnamefont
			{Ordonez-Miranda}}, \bibinfo {author} {\bibfnamefont {Y.}~\bibnamefont
			{Ezzahri}}, \bibinfo {author} {\bibfnamefont {J.~A.}\ \bibnamefont
			{Tiburcio-Moreno}}, \bibinfo {author} {\bibfnamefont {K.}~\bibnamefont
			{Joulain}},\ and\ \bibinfo {author} {\bibfnamefont {J.}~\bibnamefont
			{Drevillon}},\ }\bibfield  {title} {\bibinfo {title} {{Radiative Thermal
				Memristor}},\ }\href {https://doi.org/10.1103/PhysRevLett.123.025901}
	{\bibfield  {journal} {\bibinfo  {journal} {Phys. Rev. Lett.}\ }\textbf
		{\bibinfo {volume} {123}},\ \bibinfo {pages} {025901} (\bibinfo {year}
		{2019})}\BibitemShut {NoStop}%
	\bibitem [{\citenamefont {Ronzani}\ \emph {et~al.}(2018)\citenamefont
		{Ronzani}, \citenamefont {Karimi}, \citenamefont {Senior}, \citenamefont
		{Chang}, \citenamefont {Peltonen}, \citenamefont {Chen},\ and\ \citenamefont
		{Pekola}}]{Ronzani2018}%
	\BibitemOpen
	\bibfield  {author} {\bibinfo {author} {\bibfnamefont {A.}~\bibnamefont
			{Ronzani}}, \bibinfo {author} {\bibfnamefont {B.}~\bibnamefont {Karimi}},
		\bibinfo {author} {\bibfnamefont {J.}~\bibnamefont {Senior}}, \bibinfo
		{author} {\bibfnamefont {Y.-C.}\ \bibnamefont {Chang}}, \bibinfo {author}
		{\bibfnamefont {J.~T.}\ \bibnamefont {Peltonen}}, \bibinfo {author}
		{\bibfnamefont {C.}~\bibnamefont {Chen}},\ and\ \bibinfo {author}
		{\bibfnamefont {J.~P.}\ \bibnamefont {Pekola}},\ }\bibfield  {title}
	{\bibinfo {title} {Tunable photonic heat transport in a quantum heat valve},\
	}\href {https://doi.org/10.1038/s41567-018-0199-4} {\bibfield  {journal}
		{\bibinfo  {journal} {Nature Phys.}\ }\textbf {\bibinfo {volume} {14}},\
		\bibinfo {pages} {991} (\bibinfo {year} {2018})}\BibitemShut {NoStop}%
	\bibitem [{\citenamefont {Dutta}\ \emph {et~al.}(2020)\citenamefont {Dutta},
		\citenamefont {Majidi}, \citenamefont {Talarico}, \citenamefont {Lo~Gullo},
		\citenamefont {Courtois},\ and\ \citenamefont {Winkelmann}}]{LoGullo2020}%
	\BibitemOpen
	\bibfield  {author} {\bibinfo {author} {\bibfnamefont {B.}~\bibnamefont
			{Dutta}}, \bibinfo {author} {\bibfnamefont {D.}~\bibnamefont {Majidi}},
		\bibinfo {author} {\bibfnamefont {N.~W.}\ \bibnamefont {Talarico}}, \bibinfo
		{author} {\bibfnamefont {N.}~\bibnamefont {Lo~Gullo}}, \bibinfo {author}
		{\bibfnamefont {H.}~\bibnamefont {Courtois}},\ and\ \bibinfo {author}
		{\bibfnamefont {C.~B.}\ \bibnamefont {Winkelmann}},\ }\bibfield  {title}
	{\bibinfo {title} {{Single-Quantum-Dot Heat Valve}},\ }\href
	{https://doi.org/10.1103/PhysRevLett.125.237701} {\bibfield  {journal}
		{\bibinfo  {journal} {Phys. Rev. Lett.}\ }\textbf {\bibinfo {volume} {125}},\
		\bibinfo {pages} {237701} (\bibinfo {year} {2020})}\BibitemShut {NoStop}%
	\bibitem [{\citenamefont {Dutta}\ \emph {et~al.}(2017)\citenamefont {Dutta},
		\citenamefont {Peltonen}, \citenamefont {Antonenko}, \citenamefont {Meschke},
		\citenamefont {Skvortsov}, \citenamefont {Kubala}, \citenamefont {K\"onig},
		\citenamefont {Winkelmann}, \citenamefont {Courtois},\ and\ \citenamefont
		{Pekola}}]{Dutta2017}%
	\BibitemOpen
	\bibfield  {author} {\bibinfo {author} {\bibfnamefont {B.}~\bibnamefont
			{Dutta}}, \bibinfo {author} {\bibfnamefont {J.~T.}\ \bibnamefont {Peltonen}},
		\bibinfo {author} {\bibfnamefont {D.~S.}\ \bibnamefont {Antonenko}}, \bibinfo
		{author} {\bibfnamefont {M.}~\bibnamefont {Meschke}}, \bibinfo {author}
		{\bibfnamefont {M.~A.}\ \bibnamefont {Skvortsov}}, \bibinfo {author}
		{\bibfnamefont {B.}~\bibnamefont {Kubala}}, \bibinfo {author} {\bibfnamefont
			{J.}~\bibnamefont {K\"onig}}, \bibinfo {author} {\bibfnamefont {C.~B.}\
			\bibnamefont {Winkelmann}}, \bibinfo {author} {\bibfnamefont
			{H.}~\bibnamefont {Courtois}},\ and\ \bibinfo {author} {\bibfnamefont
			{J.~P.}\ \bibnamefont {Pekola}},\ }\bibfield  {title} {\bibinfo {title}
		{{Thermal Conductance of a Single-Electron Transistor}},\ }\href
	{https://doi.org/10.1103/PhysRevLett.119.077701} {\bibfield  {journal}
		{\bibinfo  {journal} {Phys. Rev. Lett.}\ }\textbf {\bibinfo {volume} {119}},\
		\bibinfo {pages} {077701} (\bibinfo {year} {2017})}\BibitemShut {NoStop}%
	\bibitem [{\citenamefont {Wang}\ and\ \citenamefont
		{Li}(2007)}]{wang2007thermal}%
	\BibitemOpen
	\bibfield  {author} {\bibinfo {author} {\bibfnamefont {L.}~\bibnamefont
			{Wang}}\ and\ \bibinfo {author} {\bibfnamefont {B.}~\bibnamefont {Li}},\
	}\bibfield  {title} {\bibinfo {title} {{Thermal Logic Gates: Computation with
				Phonons}},\ }\href {https://doi.org/10.1103/PhysRevLett.99.177208} {\bibfield
		{journal} {\bibinfo  {journal} {Phys. Rev. Lett.}\ }\textbf {\bibinfo
			{volume} {99}},\ \bibinfo {pages} {177208} (\bibinfo {year}
		{2007})}\BibitemShut {NoStop}%
	\bibitem [{\citenamefont {Segal}\ and\ \citenamefont
		{Nitzan}(2005)}]{Segal2005}%
	\BibitemOpen
	\bibfield  {author} {\bibinfo {author} {\bibfnamefont {D.}~\bibnamefont
			{Segal}}\ and\ \bibinfo {author} {\bibfnamefont {A.}~\bibnamefont {Nitzan}},\
	}\bibfield  {title} {\bibinfo {title} {Spin-{B}oson {T}hermal {R}ectifier},\
	}\href {https://doi.org/10.1103/PhysRevLett.94.034301} {\bibfield  {journal}
		{\bibinfo  {journal} {Phys. Rev. Lett.}\ }\textbf {\bibinfo {volume} {94}},\
		\bibinfo {pages} {034301} (\bibinfo {year} {2005})}\BibitemShut {NoStop}%
	\bibitem [{\citenamefont {Carrega}\ \emph {et~al.}(2015)\citenamefont
		{Carrega}, \citenamefont {Solinas}, \citenamefont {Braggio}, \citenamefont
		{Sassetti},\ and\ \citenamefont {Weiss}}]{Carrega2015}%
	\BibitemOpen
	\bibfield  {author} {\bibinfo {author} {\bibfnamefont {M.}~\bibnamefont
			{Carrega}}, \bibinfo {author} {\bibfnamefont {P.}~\bibnamefont {Solinas}},
		\bibinfo {author} {\bibfnamefont {A.}~\bibnamefont {Braggio}}, \bibinfo
		{author} {\bibfnamefont {M.}~\bibnamefont {Sassetti}},\ and\ \bibinfo
		{author} {\bibfnamefont {U.}~\bibnamefont {Weiss}},\ }\bibfield  {title}
	{\bibinfo {title} {Functional integral approach to time-dependent heat
			exchange in open quantum systems: general method and applications},\ }\href
	{https://doi.org/10.1088/1367-2630/17/4/045030} {\bibfield  {journal}
		{\bibinfo  {journal} {New J. Phys.}\ }\textbf {\bibinfo {volume} {17}},\
		\bibinfo {pages} {045030} (\bibinfo {year} {2015})}\BibitemShut {NoStop}%
	\bibitem [{\citenamefont {Velizhanin}\ \emph {et~al.}(2010)\citenamefont
		{Velizhanin}, \citenamefont {Thoss},\ and\ \citenamefont
		{Wang}}]{velizhanin2010meir}%
	\BibitemOpen
	\bibfield  {author} {\bibinfo {author} {\bibfnamefont {K.~A.}\ \bibnamefont
			{Velizhanin}}, \bibinfo {author} {\bibfnamefont {M.}~\bibnamefont {Thoss}},\
		and\ \bibinfo {author} {\bibfnamefont {H.}~\bibnamefont {Wang}},\ }\bibfield
	{title} {\bibinfo {title} {{Meir--Wingreen} formula for heat transport in a
			spin-boson nanojunction model},\ }\href {https://doi.org/10.1063/1.3483127}
	{\bibfield  {journal} {\bibinfo  {journal} {J. Chem. Phys.}\ }\textbf
		{\bibinfo {volume} {133}},\ \bibinfo {pages} {084503} (\bibinfo {year}
		{2010})}\BibitemShut {NoStop}%
	\bibitem [{\citenamefont {Wang}\ \emph {et~al.}(2017)\citenamefont {Wang},
		\citenamefont {Ren},\ and\ \citenamefont {Cao}}]{PhysRevA.95.023610}%
	\BibitemOpen
	\bibfield  {author} {\bibinfo {author} {\bibfnamefont {C.}~\bibnamefont
			{Wang}}, \bibinfo {author} {\bibfnamefont {J.}~\bibnamefont {Ren}},\ and\
		\bibinfo {author} {\bibfnamefont {J.}~\bibnamefont {Cao}},\ }\bibfield
	{title} {\bibinfo {title} {Unifying quantum heat transfer in a nonequilibrium
			spin-boson model with full counting statistics},\ }\href
	{https://doi.org/10.1103/PhysRevA.95.023610} {\bibfield  {journal} {\bibinfo
			{journal} {Phys. Rev. A}\ }\textbf {\bibinfo {volume} {95}},\ \bibinfo
		{pages} {023610} (\bibinfo {year} {2017})}\BibitemShut {NoStop}%
	\bibitem [{\citenamefont {Yang}\ and\ \citenamefont {Wu}(2014)}]{Yang2014}%
	\BibitemOpen
	\bibfield  {author} {\bibinfo {author} {\bibfnamefont {Y.}~\bibnamefont
			{Yang}}\ and\ \bibinfo {author} {\bibfnamefont {C.-Q.}\ \bibnamefont {Wu}},\
	}\bibfield  {title} {\bibinfo {title} {Quantum heat transport in a spin-boson
			nanojunction: Coherent and incoherent mechanisms},\ }\href
	{https://doi.org/10.1209/0295-5075/107/30003} {\bibfield  {journal} {\bibinfo
			{journal} {EPL}\ }\textbf {\bibinfo {volume} {107}},\ \bibinfo {pages}
		{30003} (\bibinfo {year} {2014})}\BibitemShut {NoStop}%
	\bibitem [{\citenamefont {Senior}\ \emph {et~al.}(2020)\citenamefont {Senior},
		\citenamefont {Gubaydullin}, \citenamefont {Karimi}, \citenamefont
		{Peltonen}, \citenamefont {Ankerhold},\ and\ \citenamefont
		{Pekola}}]{Senior2020}%
	\BibitemOpen
	\bibfield  {author} {\bibinfo {author} {\bibfnamefont {J.}~\bibnamefont
			{Senior}}, \bibinfo {author} {\bibfnamefont {A.}~\bibnamefont {Gubaydullin}},
		\bibinfo {author} {\bibfnamefont {B.}~\bibnamefont {Karimi}}, \bibinfo
		{author} {\bibfnamefont {J.~T.}\ \bibnamefont {Peltonen}}, \bibinfo {author}
		{\bibfnamefont {J.}~\bibnamefont {Ankerhold}},\ and\ \bibinfo {author}
		{\bibfnamefont {J.~P.}\ \bibnamefont {Pekola}},\ }\bibfield  {title}
	{\bibinfo {title} {Heat rectification via a superconducting artificial
			atom},\ }\href {https://doi.org/10.1038/s42005-020-0307-5} {\bibfield
		{journal} {\bibinfo  {journal} {Commun. Phys.}\ }\textbf {\bibinfo {volume}
			{3}},\ \bibinfo {pages} {40} (\bibinfo {year} {2020})}\BibitemShut {NoStop}%
	\bibitem [{\citenamefont {Belyansky}\ \emph {et~al.}()\citenamefont
		{Belyansky}, \citenamefont {Whitsitt}, \citenamefont {Lundgren},
		\citenamefont {Wang}, \citenamefont {Vrajitoarea}, \citenamefont {Houck},\
		and\ \citenamefont {Gorshkov}}]{belyansky2020transport}%
	\BibitemOpen
	\bibfield  {author} {\bibinfo {author} {\bibfnamefont {R.}~\bibnamefont
			{Belyansky}}, \bibinfo {author} {\bibfnamefont {S.}~\bibnamefont {Whitsitt}},
		\bibinfo {author} {\bibfnamefont {R.}~\bibnamefont {Lundgren}}, \bibinfo
		{author} {\bibfnamefont {Y.}~\bibnamefont {Wang}}, \bibinfo {author}
		{\bibfnamefont {A.}~\bibnamefont {Vrajitoarea}}, \bibinfo {author}
		{\bibfnamefont {A.~A.}\ \bibnamefont {Houck}},\ and\ \bibinfo {author}
		{\bibfnamefont {A.~V.}\ \bibnamefont {Gorshkov}},\ }\bibfield  {title}
	{\bibinfo {title} {Frustration-induced anomalous transport and strong photon decay in waveguide QED},\ }\href@noop {} {\bibinfo  {journal} {arXiv:2007.03690}\
	}\BibitemShut {NoStop}%
	\bibitem [{\citenamefont {Eich}\ \emph {et~al.}(2016)\citenamefont {Eich},
	\citenamefont {Ventra},\ and\ \citenamefont {Vignale}}]{Eich_2016}%
\BibitemOpen
\bibfield  {author} {\bibinfo {author} {\bibfnamefont {F.~G.}\ \bibnamefont
		{Eich}}, \bibinfo {author} {\bibfnamefont {M.~D.}\ \bibnamefont {Ventra}},\
	and\ \bibinfo {author} {\bibfnamefont {G.}~\bibnamefont {Vignale}},\
}\bibfield  {title} {\bibinfo {title} {Functional theories of thermoelectric
		phenomena},\ }\href {https://doi.org/10.1088/1361-648x/29/6/063001}
{\bibfield  {journal} {\bibinfo  {journal} {J. Phys.: Condens. Matter}\
	}\textbf {\bibinfo {volume} {29}},\ \bibinfo {pages} {063001} (\bibinfo
	{year} {2016})}\BibitemShut {NoStop}%
\bibitem [{\citenamefont {Eich}\ \emph {et~al.}(2014)\citenamefont {Eich},
	\citenamefont {Principi}, \citenamefont {Di~Ventra},\ and\ \citenamefont
	{Vignale}}]{PhysRevB.90.115116}%
\BibitemOpen
\bibfield  {author} {\bibinfo {author} {\bibfnamefont {F.~G.}\ \bibnamefont
		{Eich}}, \bibinfo {author} {\bibfnamefont {A.}~\bibnamefont {Principi}},
	\bibinfo {author} {\bibfnamefont {M.}~\bibnamefont {Di~Ventra}},\ and\
	\bibinfo {author} {\bibfnamefont {G.}~\bibnamefont {Vignale}},\ }\bibfield
{title} {\bibinfo {title} {Luttinger-field approach to thermoelectric
		transport in nanoscale conductors},\ }\href
{https://doi.org/10.1103/PhysRevB.90.115116} {\bibfield  {journal} {\bibinfo
		{journal} {Phys. Rev. B}\ }\textbf {\bibinfo {volume} {90}},\ \bibinfo
	{pages} {115116} (\bibinfo {year} {2014})}\BibitemShut {NoStop}%
\bibitem [{\citenamefont {Tatara}(2015)}]{PhysRevLett.114.196601}%
\BibitemOpen
\bibfield  {author} {\bibinfo {author} {\bibfnamefont {G.}~\bibnamefont
		{Tatara}},\ }\bibfield  {title} {\bibinfo {title} {Thermal Vector Potential
		Theory of Transport Induced by a Temperature Gradient},\ }\href
{https://doi.org/10.1103/PhysRevLett.114.196601} {\bibfield  {journal}
	{\bibinfo  {journal} {Phys. Rev. Lett.}\ }\textbf {\bibinfo {volume} {114}},\
	\bibinfo {pages} {196601} (\bibinfo {year} {2015})}\BibitemShut {NoStop}%
\bibitem [{\citenamefont {Bhandari}\ \emph {et~al.}(2020)\citenamefont
	{Bhandari}, \citenamefont {Alonso}, \citenamefont {Taddei}, \citenamefont
	{von Oppen}, \citenamefont {Fazio},\ and\ \citenamefont
	{Arrachea}}]{PhysRevB.102.155407}%
\BibitemOpen
\bibfield  {author} {\bibinfo {author} {\bibfnamefont {B.}~\bibnamefont
		{Bhandari}}, \bibinfo {author} {\bibfnamefont {P.~T.}\ \bibnamefont
		{Alonso}}, \bibinfo {author} {\bibfnamefont {F.}~\bibnamefont {Taddei}},
	\bibinfo {author} {\bibfnamefont {F.}~\bibnamefont {von Oppen}}, \bibinfo
	{author} {\bibfnamefont {R.}~\bibnamefont {Fazio}},\ and\ \bibinfo {author}
	{\bibfnamefont {L.}~\bibnamefont {Arrachea}},\ }\bibfield  {title} {\bibinfo
	{title} {Geometric properties of adiabatic quantum thermal machines},\ }\href
{https://doi.org/10.1103/PhysRevB.102.155407} {\bibfield  {journal} {\bibinfo
		{journal} {Phys. Rev. B}\ }\textbf {\bibinfo {volume} {102}},\ \bibinfo
	{pages} {155407} (\bibinfo {year} {2020})}\BibitemShut {NoStop}%
	\bibitem [{\citenamefont {Liu}\ \emph {et~al.}(2016)\citenamefont {Liu},
		\citenamefont {Xu},\ and\ \citenamefont {Wu}}]{liu2016green}%
	\BibitemOpen
	\bibfield  {journal} {  }\bibfield  {author} {\bibinfo {author} {\bibfnamefont
			{J.}~\bibnamefont {Liu}}, \bibinfo {author} {\bibfnamefont {H.}~\bibnamefont
			{Xu}},\ and\ \bibinfo {author} {\bibfnamefont {C.-Q.}\ \bibnamefont {Wu}},\
	}\bibfield  {title} {\bibinfo {title} {Green’s functions for spin boson
			systems: Beyond conventional perturbation theories},\ }\href
	{https://doi.org/10.1016/j.chemphys.2016.07.003} {\bibfield  {journal}
		{\bibinfo  {journal} {Chem. Phys.}\ }\textbf {\bibinfo {volume} {481}},\
		\bibinfo {pages} {42} (\bibinfo {year} {2016})}\BibitemShut {NoStop}%
	\bibitem [{\citenamefont {Liu}\ \emph {et~al.}(2017)\citenamefont {Liu},
		\citenamefont {Xu}, \citenamefont {Li},\ and\ \citenamefont {Wu}}]{liu2017}%
	\BibitemOpen
	\bibfield  {author} {\bibinfo {author} {\bibfnamefont {J.}~\bibnamefont
			{Liu}}, \bibinfo {author} {\bibfnamefont {H.}~\bibnamefont {Xu}}, \bibinfo
		{author} {\bibfnamefont {B.}~\bibnamefont {Li}},\ and\ \bibinfo {author}
		{\bibfnamefont {C.}~\bibnamefont {Wu}},\ }\bibfield  {title} {\bibinfo
		{title} {Energy transfer in the nonequilibrium spin-boson model: From weak to
			strong coupling},\ }\href {https://doi.org/10.1103/PhysRevE.96.012135}
	{\bibfield  {journal} {\bibinfo  {journal} {Phys. Rev. E}\ }\textbf {\bibinfo
			{volume} {96}},\ \bibinfo {pages} {012135} (\bibinfo {year}
		{2017})}\BibitemShut {NoStop}%
  \bibitem [{\citenamefont {Luttinger}(1964)}]{PhysRev.135.A1505}%
  \BibitemOpen
  \bibfield  {author} {\bibinfo {author} {\bibfnamefont {J.~M.}\ \bibnamefont
  {Luttinger}},\ }\bibfield  {title} {\bibinfo {title} {Theory of thermal
  transport coefficients},\ }\href {https://doi.org/10.1103/PhysRev.135.A1505}
  {\bibfield  {journal} {\bibinfo  {journal} {Phys. Rev.}\ }\textbf {\bibinfo
  {volume} {135}},\ \bibinfo {pages} {A1505} (\bibinfo {year}
  {1964})}\BibitemShut {NoStop}%
	\bibitem [{\citenamefont {Esposito}\ \emph {et~al.}(2015)\citenamefont
		{Esposito}, \citenamefont {Ochoa},\ and\ \citenamefont
		{Galperin}}]{esposito2015quantum}%
	\BibitemOpen
	\bibfield  {author} {\bibinfo {author} {\bibfnamefont {M.}~\bibnamefont
			{Esposito}}, \bibinfo {author} {\bibfnamefont {M.~A.}\ \bibnamefont
			{Ochoa}},\ and\ \bibinfo {author} {\bibfnamefont {M.}~\bibnamefont
			{Galperin}},\ }\bibfield  {title} {\bibinfo {title} {{Quantum Thermodynamics:
				A Nonequilibrium Green's Function Approach}},\ }\href
	{https://doi.org/10.1103/PhysRevLett.114.080602} {\bibfield  {journal}
		{\bibinfo  {journal} {Phys. Rev. Lett.}\ }\textbf {\bibinfo {volume} {114}},\
		\bibinfo {pages} {080602} (\bibinfo {year} {2015})}\BibitemShut {NoStop}%
		\bibitem [{\citenamefont {Esposito}\ \emph
  {et~al.}(2015{\natexlab{a}})\citenamefont {Esposito}, \citenamefont {Ochoa},\
  and\ \citenamefont {Galperin}}]{PhysRevB.92.235440}%
  \BibitemOpen
  \bibfield  {author} {\bibinfo {author} {\bibfnamefont {M.}~\bibnamefont
  {Esposito}}, \bibinfo {author} {\bibfnamefont {M.~A.}\ \bibnamefont
  {Ochoa}},\ and\ \bibinfo {author} {\bibfnamefont {M.}~\bibnamefont
  {Galperin}},\ }\bibfield  {title} {\bibinfo {title} {Nature of heat in
  strongly coupled open quantum systems},\ }\href
  {https://doi.org/10.1103/PhysRevB.92.235440} {\bibfield  {journal} {\bibinfo
  {journal} {Phys. Rev. B}\ }\textbf {\bibinfo {volume} {92}},\ \bibinfo
  {pages} {235440} (\bibinfo {year} {2015}{\natexlab{a}})}\BibitemShut
  {NoStop}%
	\bibitem [{\citenamefont {Ludovico}\ \emph {et~al.}(2014)\citenamefont
		{Ludovico}, \citenamefont {Lim}, \citenamefont {Moskalets}, \citenamefont
		{Arrachea},\ and\ \citenamefont {S\'anchez}}]{Ludovico2014}%
	\BibitemOpen
	\bibfield  {author} {\bibinfo {author} {\bibfnamefont {M.~F.}\ \bibnamefont
			{Ludovico}}, \bibinfo {author} {\bibfnamefont {J.~S.}\ \bibnamefont {Lim}},
		\bibinfo {author} {\bibfnamefont {M.}~\bibnamefont {Moskalets}}, \bibinfo
		{author} {\bibfnamefont {L.}~\bibnamefont {Arrachea}},\ and\ \bibinfo
		{author} {\bibfnamefont {D.}~\bibnamefont {S\'anchez}},\ }\bibfield  {title}
	{\bibinfo {title} {Dynamical energy transfer in ac-driven quantum systems},\
	}\href {https://doi.org/10.1103/PhysRevB.89.161306} {\bibfield  {journal}
		{\bibinfo  {journal} {Phys. Rev. B}\ }\textbf {\bibinfo {volume} {89}},\
		\bibinfo {pages} {161306} (\bibinfo {year} {2014})}\BibitemShut {NoStop}%
	\bibitem [{\citenamefont {Ludovico}\ \emph
		{et~al.}(2016{\natexlab{a}})\citenamefont {Ludovico}, \citenamefont
		{Moskalets}, \citenamefont {S\'anchez},\ and\ \citenamefont
		{Arrachea}}]{Ludovico2016a}%
	\BibitemOpen
	\bibfield  {author} {\bibinfo {author} {\bibfnamefont {M.~F.}\ \bibnamefont
			{Ludovico}}, \bibinfo {author} {\bibfnamefont {M.}~\bibnamefont {Moskalets}},
		\bibinfo {author} {\bibfnamefont {D.}~\bibnamefont {S\'anchez}},\ and\
		\bibinfo {author} {\bibfnamefont {L.}~\bibnamefont {Arrachea}},\ }\bibfield
	{title} {\bibinfo {title} {Dynamics of energy transport and entropy
			production in ac-driven quantum electron systems},\ }\href
	{https://doi.org/10.1103/PhysRevB.94.035436} {\bibfield  {journal} {\bibinfo
			{journal} {Phys. Rev. B}\ }\textbf {\bibinfo {volume} {94}},\ \bibinfo
		{pages} {035436} (\bibinfo {year} {2016}{\natexlab{a}})}\BibitemShut
	{NoStop}%
	\bibitem [{\citenamefont {Ludovico}\ \emph
		{et~al.}(2016{\natexlab{b}})\citenamefont {Ludovico}, \citenamefont
		{Arrachea}, \citenamefont {Moskalets},\ and\ \citenamefont
		{Sánchez}}]{Ludovico2016b}%
	\BibitemOpen
	\bibfield  {author} {\bibinfo {author} {\bibfnamefont {M.~F.}\ \bibnamefont
			{Ludovico}}, \bibinfo {author} {\bibfnamefont {L.}~\bibnamefont {Arrachea}},
		\bibinfo {author} {\bibfnamefont {M.}~\bibnamefont {Moskalets}},\ and\
		\bibinfo {author} {\bibfnamefont {D.}~\bibnamefont {Sánchez}},\ }\bibfield
	{title} {\bibinfo {title} {{Periodic Energy Transport and Entropy Production in Quantum Electronics}},\ } \href
			{https://doi.org/10.3390/e18110419}{\bibfield  {journal} {\bibinfo  {journal}
		{Entropy}\ }\textbf {\bibinfo {volume} {18}},\ \bibinfo
	{pages} {417} (\bibinfo {year}
	{2016}{\natexlab{b}})}\BibitemShut {NoStop}%
	\bibitem [{\citenamefont {Ludovico}\ \emph {et~al.}(2018)\citenamefont
		{Ludovico}, \citenamefont {Arrachea}, \citenamefont {Moskalets},\ and\
		\citenamefont {S\'anchez}}]{Ludovico2018}%
	\BibitemOpen
	\bibfield  {author} {\bibinfo {author} {\bibfnamefont {M.~F.}\ \bibnamefont
			{Ludovico}}, \bibinfo {author} {\bibfnamefont {L.}~\bibnamefont {Arrachea}},
		\bibinfo {author} {\bibfnamefont {M.}~\bibnamefont {Moskalets}},\ and\
		\bibinfo {author} {\bibfnamefont {D.}~\bibnamefont {S\'anchez}},\ }\bibfield
	{title} {\bibinfo {title} {Probing the energy reactance with adiabatically
			driven quantum dots},\ }\href {https://doi.org/10.1103/PhysRevB.97.041416}
	{\bibfield  {journal} {\bibinfo  {journal} {Phys. Rev. B}\ }\textbf {\bibinfo
			{volume} {97}},\ \bibinfo {pages} {041416} (\bibinfo {year}
		{2018})}\BibitemShut {NoStop}%
	\bibitem [{\citenamefont {Xu}\ \emph {et~al.}(2016)\citenamefont {Xu},
		\citenamefont {Wang}, \citenamefont {Zhao},\ and\ \citenamefont
		{Cao}}]{xu2016polaron}%
	\BibitemOpen
	\bibfield  {author} {\bibinfo {author} {\bibfnamefont {D.}~\bibnamefont
			{Xu}}, \bibinfo {author} {\bibfnamefont {C.}~\bibnamefont {Wang}}, \bibinfo
		{author} {\bibfnamefont {Y.}~\bibnamefont {Zhao}},\ and\ \bibinfo {author}
		{\bibfnamefont {J.}~\bibnamefont {Cao}},\ }\bibfield  {title} {\bibinfo
		{title} {Polaron effects on the performance of light-harvesting systems: a
			quantum heat engine perspective},\ }\href
	{https://doi.org/10.1088/1367-2630/18/2/023003} {\bibfield  {journal}
		{\bibinfo  {journal} {New J. Phys.}\ }\textbf {\bibinfo {volume} {18}},\
		\bibinfo {pages} {023003} (\bibinfo {year} {2016})}\BibitemShut {NoStop}%
	\bibitem [{\citenamefont {Hsieh}\ \emph {et~al.}(2019)\citenamefont {Hsieh},
		\citenamefont {Liu}, \citenamefont {Duan},\ and\ \citenamefont
		{Cao}}]{hsieh2019nonequilibrium}%
	\BibitemOpen
	\bibfield  {author} {\bibinfo {author} {\bibfnamefont {C.}~\bibnamefont
			{Hsieh}}, \bibinfo {author} {\bibfnamefont {J.}~\bibnamefont {Liu}}, \bibinfo
		{author} {\bibfnamefont {C.}~\bibnamefont {Duan}},\ and\ \bibinfo {author}
		{\bibfnamefont {J.}~\bibnamefont {Cao}},\ }\bibfield  {title} {\bibinfo
		{title} {{A Nonequilibrium Variational Polaron Theory to Study Quantum Heat
				Transport}},\ }\href {https://doi.org/10.1021/acs.jpcc.9b05607} {\bibfield
		{journal} {\bibinfo  {journal} {J. Phys. Chem. C}\ }\textbf {\bibinfo
			{volume} {123}},\ \bibinfo {pages} {17196} (\bibinfo {year}
		{2019})}\BibitemShut {NoStop}%
	\bibitem [{\citenamefont {Xu}\ and\ \citenamefont {Cao}(2016)}]{xu2016non}%
	\BibitemOpen
	\bibfield  {author} {\bibinfo {author} {\bibfnamefont {D.}~\bibnamefont
			{Xu}}\ and\ \bibinfo {author} {\bibfnamefont {J.}~\bibnamefont {Cao}},\
	}\bibfield  {title} {\bibinfo {title} {Non-canonical distribution and
			non-equilibrium transport beyond weak system-bath coupling regime: A polaron
			transformation approach},\ }\href {https://doi.org/10.1007/s11467-016-0540-2}
	{\bibfield  {journal} {\bibinfo  {journal} {Front. Phys.}\ }\textbf {\bibinfo
			{volume} {11}},\ \bibinfo {pages} {110308} (\bibinfo {year}
		{2016})}\BibitemShut {NoStop}%
	\bibitem [{\citenamefont {Mao}\ \emph {et~al.}(2003)\citenamefont {Mao},
		\citenamefont {Coleman}, \citenamefont {Hooley},\ and\ \citenamefont
		{Langreth}}]{PhysRevLett.91.207203}%
	\BibitemOpen
	\bibfield  {author} {\bibinfo {author} {\bibfnamefont {W.}~\bibnamefont
			{Mao}}, \bibinfo {author} {\bibfnamefont {P.}~\bibnamefont {Coleman}},
		\bibinfo {author} {\bibfnamefont {C.}~\bibnamefont {Hooley}},\ and\ \bibinfo
		{author} {\bibfnamefont {D.}~\bibnamefont {Langreth}},\ }\bibfield  {title}
	{\bibinfo {title} {{Spin Dynamics from Majorana Fermions}},\ }\href
	{https://doi.org/10.1103/PhysRevLett.91.207203} {\bibfield  {journal}
		{\bibinfo  {journal} {Phys. Rev. Lett.}\ }\textbf {\bibinfo {volume} {91}},\
		\bibinfo {pages} {207203} (\bibinfo {year} {2003})}\BibitemShut {NoStop}%
	\bibitem [{\citenamefont {Schad}\ \emph {et~al.}(2016)\citenamefont {Schad},
		\citenamefont {Shnirman},\ and\ \citenamefont
		{Makhlin}}]{PhysRevB.93.174420}%
	\BibitemOpen
	\bibfield  {author} {\bibinfo {author} {\bibfnamefont {P.}~\bibnamefont
			{Schad}}, \bibinfo {author} {\bibfnamefont {A.}~\bibnamefont {Shnirman}},\
		and\ \bibinfo {author} {\bibfnamefont {Y.}~\bibnamefont {Makhlin}},\
	}\bibfield  {title} {\bibinfo {title} {Using {M}ajorana spin-$\frac{1}{2}$
			representation for the spin-boson model},\ }\href
	{https://doi.org/10.1103/PhysRevB.93.174420} {\bibfield  {journal} {\bibinfo
			{journal} {Phys. Rev. B}\ }\textbf {\bibinfo {volume} {93}},\ \bibinfo
		{pages} {174420} (\bibinfo {year} {2016})}\BibitemShut {NoStop}%
	\bibitem [{\citenamefont {Agarwalla}\ and\ \citenamefont
		{Segal}(2017)}]{agarwalla2017}%
	\BibitemOpen
	\bibfield  {author} {\bibinfo {author} {\bibfnamefont {B.~K.}\ \bibnamefont
			{Agarwalla}}\ and\ \bibinfo {author} {\bibfnamefont {D.}~\bibnamefont
			{Segal}},\ }\bibfield  {title} {\bibinfo {title} {Energy current and its
			statistics in the nonequilibrium spin-boson model: {M}ajorana fermion
			representation},\ }\href {https://doi.org/10.1088/1367-2630/aa6657}
	{\bibfield  {journal} {\bibinfo  {journal} {New J. Phys.}\ }\textbf {\bibinfo
			{volume} {19}},\ \bibinfo {pages} {043030} (\bibinfo {year}
		{2017})}\BibitemShut {NoStop}%
	\bibitem [{\citenamefont {Keldysh}(1965)}]{keldysh1965diagram}%
	\BibitemOpen
	\bibfield  {author} {\bibinfo {author} {\bibfnamefont {L.~V.}\ \bibnamefont
			{Keldysh}},\ }\bibfield  {title} {\bibinfo {title} {Diagram technique for
			nonequilibrium processes},\ }\href@noop {} {\bibfield  {journal} {\bibinfo
			{journal} {Sov. Phys. JETP}\ }\textbf {\bibinfo {volume} {20}},\ \bibinfo
		{pages} {1018} (\bibinfo {year} {1965})}\BibitemShut {NoStop}%
	\bibitem [{\citenamefont {Haug}\ and\ \citenamefont {Jauho}(2008)}]{Haug2008}%
	\BibitemOpen
	\bibfield  {author} {\bibinfo {author} {\bibfnamefont {H.}~\bibnamefont
			{Haug}}\ and\ \bibinfo {author} {\bibfnamefont {A.-P.}\ \bibnamefont
			{Jauho}},\ }\href@noop {} {\emph {\bibinfo {title} {Quantum Kinetics in
				Transport and Optics of Semiconductors}}}\ (\bibinfo  {publisher}
	{Springer},\ \bibinfo {year} {2008})\BibitemShut {NoStop}%
	\bibitem [{\citenamefont {Talarico}\ \emph {et~al.}(2019)\citenamefont
		{Talarico}, \citenamefont {Maniscalco},\ and\ \citenamefont
		{Gullo}}]{Talarico2019}%
	\BibitemOpen
	\bibfield  {author} {\bibinfo {author} {\bibfnamefont {N.~W.}\ \bibnamefont
			{Talarico}}, \bibinfo {author} {\bibfnamefont {S.}~\bibnamefont
			{Maniscalco}},\ and\ \bibinfo {author} {\bibfnamefont {N.~L.}\ \bibnamefont
			{Gullo}},\ }\bibfield  {title} {\bibinfo {title} {{A Scalable Numerical
				Approach to the Solution of the Dyson Equation for the Non-Equilibrium
				Single-Particle Green's Function}},\ }\href
	{https://doi.org/10.1002/pssb.201800501} {\bibfield  {journal} {\bibinfo
			{journal} {Phys. Status Solidi B}\ }\textbf {\bibinfo {volume} {256}},\
		\bibinfo {pages} {1800501} (\bibinfo {year} {2019})}\BibitemShut {NoStop}%
	\bibitem [{\citenamefont {Talarico}\ \emph {et~al.}(2020)\citenamefont
		{Talarico}, \citenamefont {Maniscalco},\ and\ \citenamefont
		{Gullo}}]{Talarico2020}%
	\BibitemOpen
	\bibfield  {author} {\bibinfo {author} {\bibfnamefont {N.~W.}\ \bibnamefont
			{Talarico}}, \bibinfo {author} {\bibfnamefont {S.}~\bibnamefont
			{Maniscalco}},\ and\ \bibinfo {author} {\bibfnamefont {N.~L.}\ \bibnamefont
			{Gullo}},\ }\bibfield  {title} {\bibinfo {title} {Study of the energy
			variation in many-body open quantum systems: Role of interactions in the weak
			and strong coupling regimes},\ }\href
	{https://doi.org/10.1103/PhysRevB.101.045103} {\bibfield  {journal} {\bibinfo
			{journal} {Phys. Rev. B}\ }\textbf {\bibinfo {volume} {101}},\ \bibinfo
		{pages} {045103} (\bibinfo {year} {2020})}\BibitemShut {NoStop}%
	\bibitem [{\citenamefont {Weiss}(2012)}]{weiss2012quantum}%
	\BibitemOpen
	\bibfield  {author} {\bibinfo {author} {\bibfnamefont {U.}~\bibnamefont
			{Weiss}},\ }\href@noop {} {\emph {\bibinfo {title} {Quantum dissipative
				systems}}},\ Vol.~\bibinfo {volume} {13}\ (\bibinfo  {publisher} {World
		scientific},\ \bibinfo {year} {2012})\BibitemShut {NoStop}%
	\bibitem [{\citenamefont {Dattagupta}\ \emph {et~al.}(1989)\citenamefont
		{Dattagupta}, \citenamefont {Grabert},\ and\ \citenamefont
		{Jung}}]{Dattagupta_1989}%
	\BibitemOpen
	\bibfield  {author} {\bibinfo {author} {\bibfnamefont {S.}~\bibnamefont
			{Dattagupta}}, \bibinfo {author} {\bibfnamefont {H.}~\bibnamefont
			{Grabert}},\ and\ \bibinfo {author} {\bibfnamefont {R.}~\bibnamefont
			{Jung}},\ }\bibfield  {title} {\bibinfo {title} {The structure factor for
			neutron scattering from a two-state system in metals},\ }\href
	{https://doi.org/10.1088/0953-8984/1/8/003} {\bibfield  {journal} {\bibinfo
			{journal} {J. Phys.: Condens. Matter}\ }\textbf {\bibinfo {volume} {1}},\
		\bibinfo {pages} {1405} (\bibinfo {year} {1989})}\BibitemShut {NoStop}%
	\bibitem [{\citenamefont {Krantz}\ \emph {et~al.}(2019)\citenamefont {Krantz},
		\citenamefont {Kjaergaard}, \citenamefont {Yan}, \citenamefont {Orlando},
		\citenamefont {Gustavsson},\ and\ \citenamefont {Oliver}}]{Krantz2019}%
	\BibitemOpen
	\bibfield  {author} {\bibinfo {author} {\bibfnamefont {P.}~\bibnamefont
			{Krantz}}, \bibinfo {author} {\bibfnamefont {M.}~\bibnamefont {Kjaergaard}},
		\bibinfo {author} {\bibfnamefont {F.}~\bibnamefont {Yan}}, \bibinfo {author}
		{\bibfnamefont {T.~P.}\ \bibnamefont {Orlando}}, \bibinfo {author}
		{\bibfnamefont {S.}~\bibnamefont {Gustavsson}},\ and\ \bibinfo {author}
		{\bibfnamefont {W.~D.}\ \bibnamefont {Oliver}},\ }\bibfield  {title}
	{\bibinfo {title} {A quantum engineer's guide to superconducting qubits},\
	}\href {https://doi.org/10.1063/1.5089550} {\bibfield  {journal} {\bibinfo
			{journal} {Appl. Phys. Rev.}\ }\textbf {\bibinfo {volume} {6}},\ \bibinfo
		{pages} {021318} (\bibinfo {year} {2019})}\BibitemShut {NoStop}%
	\bibitem [{\citenamefont {Brantut}\ \emph {et~al.}(2013)\citenamefont
		{Brantut}, \citenamefont {Grenier}, \citenamefont {Meineke}, \citenamefont
		{Stadler}, \citenamefont {Krinner}, \citenamefont {Kollath}, \citenamefont
		{Esslinger},\ and\ \citenamefont {Georges}}]{Brantut2013}%
	\BibitemOpen
	\bibfield  {author} {\bibinfo {author} {\bibfnamefont {J.-P.}\ \bibnamefont
			{Brantut}}, \bibinfo {author} {\bibfnamefont {C.}~\bibnamefont {Grenier}},
		\bibinfo {author} {\bibfnamefont {J.}~\bibnamefont {Meineke}}, \bibinfo
		{author} {\bibfnamefont {D.}~\bibnamefont {Stadler}}, \bibinfo {author}
		{\bibfnamefont {S.}~\bibnamefont {Krinner}}, \bibinfo {author} {\bibfnamefont
			{C.}~\bibnamefont {Kollath}}, \bibinfo {author} {\bibfnamefont
			{T.}~\bibnamefont {Esslinger}},\ and\ \bibinfo {author} {\bibfnamefont
			{A.}~\bibnamefont {Georges}},\ }\bibfield  {title} {\bibinfo {title} {{A
				Thermoelectric Heat Engine with Ultracold Atoms}},\ }\href
	{https://doi.org/10.1126/science.1242308} {\bibfield  {journal} {\bibinfo
			{journal} {Science}\ }\textbf {\bibinfo {volume} {342}},\ \bibinfo {pages}
		{713} (\bibinfo {year} {2013})}\BibitemShut {NoStop}%
	\bibitem [{\citenamefont {Ro{\ss}nagel}\ \emph {et~al.}(2016)\citenamefont
		{Ro{\ss}nagel}, \citenamefont {Dawkins}, \citenamefont {Tolazzi},
		\citenamefont {Abah}, \citenamefont {Lutz}, \citenamefont {Schmidt-Kaler},\
		and\ \citenamefont {Singer}}]{Rossnagel2016}%
	\BibitemOpen
	\bibfield  {author} {\bibinfo {author} {\bibfnamefont {J.}~\bibnamefont
			{Ro{\ss}nagel}}, \bibinfo {author} {\bibfnamefont {S.~T.}\ \bibnamefont
			{Dawkins}}, \bibinfo {author} {\bibfnamefont {K.~N.}\ \bibnamefont
			{Tolazzi}}, \bibinfo {author} {\bibfnamefont {O.}~\bibnamefont {Abah}},
		\bibinfo {author} {\bibfnamefont {E.}~\bibnamefont {Lutz}}, \bibinfo {author}
		{\bibfnamefont {F.}~\bibnamefont {Schmidt-Kaler}},\ and\ \bibinfo {author}
		{\bibfnamefont {K.}~\bibnamefont {Singer}},\ }\bibfield  {title} {\bibinfo
		{title} {A single-atom heat engine},\ }\href
	{https://doi.org/10.1126/science.aad6320} {\bibfield  {journal} {\bibinfo
			{journal} {Science}\ }\textbf {\bibinfo {volume} {352}},\ \bibinfo {pages}
		{325} (\bibinfo {year} {2016})}\BibitemShut {NoStop}%
	\bibitem [{\citenamefont {Hanson}\ \emph {et~al.}(2007)\citenamefont {Hanson},
		\citenamefont {Kouwenhoven}, \citenamefont {Petta}, \citenamefont {Tarucha},\
		and\ \citenamefont {Vandersypen}}]{Hanson2007}%
	\BibitemOpen
	\bibfield  {author} {\bibinfo {author} {\bibfnamefont {R.}~\bibnamefont
			{Hanson}}, \bibinfo {author} {\bibfnamefont {L.~P.}\ \bibnamefont
			{Kouwenhoven}}, \bibinfo {author} {\bibfnamefont {J.~R.}\ \bibnamefont
			{Petta}}, \bibinfo {author} {\bibfnamefont {S.}~\bibnamefont {Tarucha}},\
		and\ \bibinfo {author} {\bibfnamefont {L.~M.~K.}\ \bibnamefont
			{Vandersypen}},\ }\bibfield  {title} {\bibinfo {title} {Spins in few-electron
			quantum dots},\ }\href {https://doi.org/10.1103/RevModPhys.79.1217}
	{\bibfield  {journal} {\bibinfo  {journal} {Rev. Mod. Phys.}\ }\textbf
		{\bibinfo {volume} {79}},\ \bibinfo {pages} {1217} (\bibinfo {year}
		{2007})}\BibitemShut {NoStop}%
\end{thebibliography}%
\end{document}